\documentclass[preprint,5p,twocolumn,sort&compress,times,fleqn]{elsarticle}

\pdfoutput=1
\usepackage[unicode=true, bookmarks=true,bookmarksnumbered=true, bookmarksopen=true, breaklinks=false, backref=true, pagebackref=false, colorlinks=true] { hyperref}
\hypersetup{pdftitle={SOSpin}, pdfauthor={David Costa, Nuno Cardoso, Nuno Goncalves and Catarina Simoes}, linkcolor=black, citecolor=black, urlcolor=black, filecolor=black, pdfpagelayout=TwoColumn, pdfnewwindow=true, pdfstartview=XYZ, plainpages=false}

\usepackage[utf8]{inputenc}
\usepackage[T1]{fontenc}
\usepackage{amsmath,amsthm,amssymb,amsfonts,bbm}
\usepackage{color}
\usepackage{xspace}
\usepackage{mathrsfs}
\usepackage{graphicx}
\usepackage{braket}
\usepackage{url}
\usepackage[makeroom]{cancel}

\usepackage{cleveref}
\crefname{section}{Section}{Sections}
\Crefname{section}{Section}{Sections}
\crefname{equation}{Eq.}{Eqs.}
\Crefname{equation}{Eq.}{Eqs.}
\crefname{figure}{Fig.}{Figs.}
\Crefname{figure}{Fig.}{Figs.}
\crefname{table}{Table}{Tables}
\Crefname{table}{Table}{Tables}
\crefname{appendix}{}{}
\Crefname{appendix}{}{}

\definecolor{violet}{RGB}{127, 0, 255}
\definecolor{cobalto}{RGB}{0, 40, 100}
\definecolor{mygreen}{rgb}{0,0.6,0}
\definecolor{mygray}{rgb}{0.5,0.5,0.5}
\definecolor{mymauve}{rgb}{0.58,0,0.82}
\definecolor{deepblue}{rgb}{0,0,0.5}
\definecolor{deepred}{rgb}{0.6,0,0}
\definecolor{deepgreen}{rgb}{0,0.5,0}
\definecolor{lblue}{rgb}{0.25,0.35,0.75}

\newcommand{\SOSpin}{
\mbox{\!{\color{cobalto} S}\hspace{-0.5mm}{\color{cobalto}O}\hspace{-0.39mm}{\color{black}{\scalebox{-1}[1]{S}}}{\color{black}\hspace{-0.6mm}p}{\color{black}\hspace{-0.6mm}i}{\color{black}\hspace{-0.6mm}n}}\xspace}

\usepackage{listings}
\newcommand{\code}[1]{\lstinline[basicstyle=\normalsize\ttfamily,breaklines=true]!#1!}
\newcommand{\codeb}[1]{\lstinline[basicstyle=\normalsize\ttfamily,deleteemph={b,bt},breaklines=true]!#1!}

\newcommand*{\Cpp}{C\ensuremath{++}\xspace}

\DeclareMathOperator{\ord}{Ord}
\DeclareMathOperator{\tr}{Tr}

\journal{Computer Physics Communications}
 
\begin{document}

\begin{frontmatter}

\title{\protect\SOSpin, a \Cpp library for Yukawa decomposition in $\mathsf{SO}(2N)$ models}

\author[USA,lisboa]{Nuno Cardoso}
\ead{nuno.cardoso@tecnico.ulisboa.pt}

\author[lisboa]{David Emmanuel-Costa\corref{cor1}}
\ead{david.costa@tecnico.ulisboa.pt}

\author[coimbra]{Nuno Gon\c{c}alves}
\ead{nunogon@deec.uc.pt}

\author[liege]{C.~Sim\~oes}
\ead{csimoes@ulg.ac.be}

\cortext[cor1]{Corresponding author}

\address[USA]{NCSA,
University of Illinois at Urbana-Champaign,
1205 W. Clark St.,
Urbana, IL 61801, 
USA}

\address[lisboa]{CFTP, Departamento de F\'{i}sica, 
Instituto Superior T\'{e}cnico, 
Universidade de Lisboa, 
Av. Rovisco Pais, 
1049-001 Lisbon, 
Portugal}

\address[coimbra]{Institute of Systems and Robotics, 
Department of Electrical and Computer Engineering, 
University of Coimbra, 
Polo 2 - Pinhal de Marrocos, 
3030-290 Coimbra, 
Portugal}

\address[liege]{IFPA, D\'ep. AGO, 
Quartier Agora, 19A All\'ee du 6 ao\^ut, B\^at B5a, 
Universit\'e de Li\`ege, 4000 Li\`ege, 
Belgique}

\begin{abstract}
We present in this paper the \protect\SOSpin library, which calculates an analytic decomposition of the Yukawa interactions invariant under $\mathsf{SO}(2N)$ in terms of an $\mathsf{SU}(N)$ basis. We make use of the oscillator expansion formalism, where the $\mathsf{SO}(2N)$ spinor representations are expressed in terms of creation and annihilation operators of a Grassmann algebra acting on a vacuum state. These noncommutative operators and their products are simulated in \protect\SOSpin through the implementation of doubly-linked-list data structures. These data structures were determinant to achieve a higher performance in the simplification of large products of creation and annihilation operators. We illustrate the use of our library with complete examples of how to decompose Yukawa terms invariant under $\mathsf{SO}(2N)$ in terms of $\mathsf{SU}(N)$ degrees of freedom for $N=2$ and $5$. We further demonstrate, with an example for $\mathsf{SO}(4)$, that higher dimensional field-operator terms can also be processed with our library. Finally, we describe the functions available in \protect\SOSpin that are made to simplify the writing of spinors and their interactions specifically for $\mathsf{SO}(10)$ models.
\end{abstract}

\begin{keyword}
Special orthogonal groups \sep Grand Unified Theory

\PACS 02.20.Qs \sep 02.70.Wz \sep 12.10.-g \sep 12.10.Dm 

\emph{Prepint}: CFTP/15-008

Version: 1.0 
\end{keyword}
\end{frontmatter}

\tableofcontents\vspace*{\baselineskip}

%%%%%%%%%%%%%%%%%%%%%%%%%%%%%%%%%%%%%%
\section{Introduction}

The orthogonal groups $\mathsf{O}(N)$ and their generalisations have played an important role in the construction of modern physics. In particular, the special orthogonal groups $\mathsf{SO}(N)$ appear naturally in the context of physical systems invariant under rotations, which in turn implies the conservation of the angular momentum or the determination of the azimuthal quantum number for an atomic orbital. The notion of spin used to describe the intrinsic angular momentum of particles is another example of the importance of special orthogonal groups. Indeed, the spin group $\mathsf{Spin}(N)$ is a double cover of the special orthogonal group $\mathsf{SO}(N)$, i.e., $\mathsf{Spin}(N)$ is locally isomorphic to $\mathsf{SO}(N)$ (see, e.g., Ref.~\cite{Baez:2009dj}).

In particle physics, the use of special orthogonal groups $\mathsf{SO}(N)$ have been very productive in the construction of Grand Unified Theories (GUTs). The original idea of GUT models is to embed the Standard Model~(SM) gauge group $\mathsf{SU}(3)_{\text{\sc c}}\times \mathsf{SU}(2)_{\text{\sc l}}\times \mathsf{U}(1)_{\text{\sc y}}$ in a larger simple Lie group, so that the three SM gauge couplings unify into a unique coupling. The first GUT model was proposed by Georgi and Glashow~\cite{Georgi:1974sy} in 1974 and it introduced $\mathsf{SU}(5)$ as the unifying gauge group. The group $\mathsf{SU}(5)$ has rank 4 as the SM group and the observed fermions are grouped in two unique representations $\bar{\mathsf{5}}$ and $\mathsf{10}$, per generation. 

The possibility of having a GUT model based on the special orthogonal group $\mathsf{SO}(10)$ was first accounted by Georgi~\cite{Georgi:1974my,Georgi:1975qb} and Fritzsch and Minkowski~\cite{Fritzsch:1974nn}. The $\mathsf{SO}(10)$ model brought new interesting features over $\mathsf{SU}(5)$. Each generation of SM fermions are accommodated in a unique $\mathsf{16}$-spinorial representation of $\mathsf{SO}(10)$ with an additional place for a singlet Weyl field, that can be interpreted later as a right-handed neutrino. These sterile neutrino states allow naturally to explain the observed oscillations of neutrinos through the Seesaw mechanism~\cite{Minkowski:1977sc,Yanagida:1979as,Mohapatra:1979ia,Schechter:1980gr,GellMann:1980vs}; giving an extremely light mass to the active neutrinos when the sterile neutrino mass is of order of the unification scale. The $\mathsf{SO}(10)$ gauge interactions conserve parity thus making parity a continuous symmetry. Due to the fact that the rank of $\mathsf{SO}(10)$ is 5, there is an extra diagonal generator with quantum number $B-L$ as in the left-right symmetric models and it is indeed the minimal left-right symmetric GUT model. Finally, GUT models based on $\mathsf{SO}(N)$, apart from $\mathsf{SO}(6)$, turn out to be automatically free of gauge anomalies~\cite{Georgi:1972bb}. 

Since the appearance of first $\mathsf{SO}(10)$ GUT model, many models based on $\mathsf{SO}(10)$ have been proposed in the literature (cf. Refs.~\cite{Harvey:1980je,Rajpoot:1980xy,Harvey:1981hk,Wilczek:1981iz,Barr:1981qv,Babu:1998wi,Bertolini:2009qj,Drees:2008tc,Fong:2014gea,Fonseca:2015aoa,Babu:2015bna} and references therein). In addition, other models were implemented within $\mathsf{SO}(N)$ unification with a rank greater than 5, e.g., $\mathsf{SO}(12)$ ~\cite{Rajpoot:1981it}, $\mathsf{SO}(14)$~\cite{Ida:1980ea}, and $\mathsf{SO}(18)$~\cite{Fujimoto:1981bv,Chang:1985uf,Hubsch:1985zn}. $\mathsf{SO}(18)$ turns out to be the minimal special orthogonal group that accommodates the three SM fermionic generations in a unique spinorial representation $\mathsf{256}$ by choosing properly the breaking chain down to the SM. There are also applications of $\mathsf{SO}(N)$ as unifying group in the context of models with extra-dimensions, e.g., $\mathsf{SO}(10)$ in 5D~\cite{Kim:2002im,Feruglio:2014jla}, in orbifold 5D~\cite{Hebecker:2001jb} and 6D~\cite{Hebecker:2001jb,Asaka:2001ez,Asaka:2001eh,Asaka:2002nd,Asaka:2003iy,Buchmuller:2004eg,Buchmuller:2007xv}. The group $\mathsf{SO}(11)$ was also used in the context of Randall-Sundrum warped space~\cite{Cosme:2003cq,Hosotani:2015hoa}.
 
The breaking of a GUT $\mathsf{SO}(N)$ model down to the SM can be achieved by different breaking path, with possibly some intermediate mass scales. In order to understand the possible $\mathsf{SO}(N)$ breaking paths, it is important to identify its maximal subgroup (with the same rank as the higher group), so that one can express representations of $\mathsf{SO}(N)$ in terms of representations of the maximal subgroup and therefore understand the necessary Higgs sector. In particular, for the group $\mathsf{SO}(10)$ one identifies two important maximal subgroups~\cite{Rajpoot:1980xy,Barr:1981qv}, namely $\mathsf{SU}(5)\times\mathsf{U}(1)$ and $\mathsf{SO}(6)\times\mathsf{SO}(4)$, which is equivalent to $\mathsf{SU}(4)\times \mathsf{SU}(2)_{\text{\sc l}}\times \mathsf{SU}(2)_{\text{\sc r}}$. The first subgroup can be broken into the usual $\mathsf{SU}(5)$. Instead, the second subgroup can be broken into the Pati-Salam model, $\mathsf{SU}(3)\times \mathsf{SU}(2)_{\text{\sc l}}\times \mathsf{SU}(2)_{\text{\sc r}}\times \mathsf{U}(1)_{B-L}$, in which the $B-L$ symmetry of the SM is gauged. It is worth to point out that one can also break $\mathsf{SO}(10)$ to the flipped-$\mathsf{SU}(5)$~\cite{Barr:1981qv}, where the SM hypercharge is identified with a linear combination of the diagonal generator of $\mathsf{SU}(5)$ with extra $\mathsf{U}(1)$ generator of $\mathsf{SO}(10)$.

The purpose of this paper is to introduce the \SOSpin library implemented in the \Cpp programming language. The idea behind the conception of \SOSpin is the decomposition of Yukawa interactions invariant under $\mathsf{SO}(2N)$ in terms of $\mathsf{SU}(N)$ degrees of freedom. This decomposition is particularly useful for GUT models based on $\mathsf{SO}(2N)$ that break to an intermediate threshold symmetric under $\mathsf{SU}(N)$, since it allows to relate the Yukawa couplings in the intermediate theory with the GUT Yukawa couplings from the GUT theory, and thus leading to predictions. In general, this decomposition can be fastidious and error-prone. Our library is meant to simplify this task.

The \SOSpin library relies on the oscillator expansion formalism, where the $\mathsf{SO}(N)$ spinor are written in an $\mathsf{SU}(N)$ basis realised through the introduction of creation and annihilation operators of a Grassmann algebra~\cite{Mohapatra:1979nn}. These operators and their algebra are simulated in \SOSpin by means of doubly-linked lists as the appropriate data structure for these problems. This type of data structure has higher performance power, since it optimises the memory usage for long chains of operators and the data itself in memory do not need to be adjacent. Although the \SOSpin library was projected with the groups $\mathsf{SO}(2N)$ in mind, it can be easily adapted to the groups $\mathsf{SO}(2N+1)$ or even to other systems where creation and annihilation operators can be defined.

The paper is organised as follows. In the next section, we discuss the spinorial representations of $\mathsf{SO}(2N)$ in a basis in terms of the degrees of freedom of $\mathsf{SU}(N)$, through creation and annihilation operators defined in a Grassmann algebra. We then apply this method to decompose Yukawa interactions invariant under $\mathsf{SO}(2N)$ in terms of $\mathsf{SU}(N)$ interactions. In \cref{sec:prog}, we present the general structure of the \SOSpin library, giving in detail the general functions and specific functions for $\mathsf{SO}(10)$. In \cref{sec:workprog}, we explain the installation of our library and we show how to write simple programs. Then in \cref{sec:examples}, we give complete examples for computing Yukawa terms in $\mathsf{SO}(4)$ and $\mathsf{SO}(10)$ with the \SOSpin library. Finally, we draw our conclusions in \cref{sec:conclusions}. 

%%%%%%%%%%%%%%%%%%%%%%%%%%%%%%%%%%%%%%
\section{The $\mathbf{SO}(2N)$ spinor representation}
\label{sec:sec1}

We review in this section the oscillator expansion technique~\cite{Mohapatra:1979nn, Wilczek:1981iz,Nandi:1981py} that is implemented in the \SOSpin library. This technique has been actively explored for explicit computations of $\mathsf{SO}(10)$ Yukawa couplings~\cite{Nath:2001uw,Nath:2001yj,Nath:2005bx}. The main idea of this technique is to write the two spinor representations of $\mathsf{SO}(2N)$ in a basis where the spinor components are expressed explicitly in terms of $\mathsf{SU}(N)$ fields. This is achieved by constructing a Grassmann algebra of creation and annihilation operators. One could have used a completely group theoretical approach as done in Ref.~\cite{Anderson:2001sd}, but the oscillator expansion technique is more field theoretical and seems more intuitive to consider the case where the breaking of $\mathsf{SO}(2N)$ is done down to $\mathsf{SU}(N)$. In addition there are other methods in the literature~\cite{He:1990jw,Fukuyama:2004ps,Aulakh:2002zr,Aulakh:2005sq} that can be used for computing the $\mathsf{SO}(2N)$ invariant couplings, but we shall not consider these methods in this paper.

We start by introducing the general properties of any special orthogonal group $\mathsf{SO}(N)$, which are the simple Lie group of all orthogonal $N\times N$ matrices $\mathbf{O}$ such that
\begin{equation}
\mathbf{O}^{\intercal}\mathbf{O}\,=\,\mathbf{O}\,\mathbf{O}^{\intercal}=\,\mathbf{1}\,,
\end{equation}
with the special condition $|\mathbf{O}|=1$. This group leaves invariant the bilinear 
\begin{equation}
\mathbf{x}^{\intercal}\,\mathbf{y} \,=\,
x_1\,y_1 \,+\, x_2\,y_2 \,+\, \cdots \,+\, x_N\,y_N\,,
\end{equation}
when the N-dimensional vectors $\mathbf{x}$ and $\mathbf{y}$ transform as 
\begin{equation}
\label{eq:genSO}
x_{\mu}\longrightarrow x^{\prime}_{\mu} = O_{\mu\nu}\, x_{\nu}\,,
\quad
y_{\mu} \longrightarrow y^{\prime}_{\mu} = O_{\mu\nu}\, y_{\nu}\,. 
\end{equation}
Making an infinitesimal group transformation, the matrix elements $O_{\mu\nu}$ can be expanded as
\begin{equation}
\label{eq:O}
O_{\mu\nu}\,=\, \delta_{\mu\nu} \,-\, \frac{i}{2}\,\omega_{\mu\nu}M_{\mu\nu} \,+\, \mathcal{O}(\omega^2)\,,
\end{equation}
where $\omega_{\mu\nu}$ is a real antisymmetric tensor, while $M_{\mu\nu}$ are $N(N-1)/2$ independent $N\times N$-matrix generators of $\mathsf{SO}(N)$. In the vector representation, the generators are hermitian, $M_{\mu\nu}^{\dagger}=M_{\mu\nu}$, and they can be written as
\begin{equation}
\left(M_{\mu\nu}\right)_{mn}\,=\, i\left(\delta_{\mu m}\delta_{\nu n}-\delta_{\nu m}\delta_{\mu n}\right)\,,
\end{equation}
implying $\tr M_{\mu\nu}=0$, and they satisfy the Lie algebra of $\mathsf{SO}(N)$ as
\begin{equation}
\label{eq:algebra}
\begin{split}
\left[M_{\mu\nu},\,M_{\rho\eta}\right] = i\left(
\delta_{\mu\eta}M_{\nu\rho} - \delta_{\mu\rho}M_{\nu\eta}
-\delta_{\nu\eta}M_{\mu\rho} + \delta_{\nu\rho}M_{\mu\eta}
\right).
\end{split}
\end{equation}
Within the Cartan classification, the Lie algebra associated to the group $\mathsf{SO}(2N+1)$ is $B_{N}$ while to $\mathsf{SO}(2N)$ is $D_N$, with $N$ being identified as the rank of the algebra. We focus now our discussion only on even-dimensional special groups $\mathsf{SO}(2N)$. Note that the oscillator expansion technique can also be applied to the spinor representation of $\mathsf{SO}(2N+1)$. 

The spinor representations of $\mathsf{SO}(2N)$ can be constructed if one introduces a set of matrices $\{\Gamma_{\mu}\}$, with $\mu=1,\dots N$, such that 
\begin{equation}
\label{eq:inv2}
x_1\,\Gamma^2_1\,y_1 \,+\, x_2\,\Gamma^2_2\,y_2 \,+\, \cdots \,+\, x_N\,\Gamma^2_N\,y_N\,=\,\mathbf{x}^{\intercal}\,\mathbf{y}\,.
\end{equation}
In order to verify \cref{eq:inv2}, one must necessarily impose that the matrices $\Gamma_{\mu}$ should obey to:
\begin{equation}
\{\Gamma_{\mu},\Gamma_{\nu}\}=2\delta_{\mu \nu}\,,
\end{equation}
which form a Clifford algebra. It is straightforward to see that any ordered product of distinct gamma matrices gives rise to a complete set of linearly independent matrices. This fact leads to the construction of the so-called spinor representation of $\mathsf{SO}(2N)$. In \cref{sec:clifford} we give a general proof of the existence of the matrices $\Gamma_{\mu}$. In fact, for any even-dimensional Clifford algebra there is only one irreducible representation of dimension $2^N$. Instead of writing explicitly the matrices $\Gamma_{\mu}$ via the $2^N\times 2^N$ generalised Dirac matrices formed from the direct product of the Pauli matrices, we write them in terms of a set of creation ($b^{\dagger}_i$) and annihilation ($b_i$) operators acting on the Hilbert space as
\begin{equation}
\label{eq:anticomrelation}
\{b_i,b^{\dagger}_j\}=\delta_{ij}\,,\quad\{b_i,b_j\}=0=\{b^{\dagger}_i,b^{\dagger}_j\}\,,
\end{equation}
with $i=1,...,N$. Each pair $b_i$, $b^{\dagger}_i$ of operators can be constructed directly from linear combinations of pairs of $\Gamma$-matrices as
\begin{equation}
\label{eq:gammamatrices}
b_j=\frac12(i\, \Gamma_{2j-1} \,+\, \Gamma_{2j})\,,\quad b^{\dagger}_j=\frac12(-i\, \Gamma_{2j-1} \,+\, \Gamma_{2j}),
\end{equation}
with the inverted relation given by
\begin{equation}
\label{eq:gammamatricesrev}
\Gamma_{2j-1}=-i(b_j-b^{\dagger}_j)\,,\quad \Gamma_{2j}=(b_j+b^{\dagger}_j),
\end{equation}
showing a one-to-one correspondence. General formulae for the correspondence between the Clifford and the Grassmann algebrae are found in \cref{sec:CliffvsGrass}. 

The advantage of this approach is that one does not need to write explicitly the operators $b_i$, $b^{\dagger}_i$, one needs only to define the vacuum state $\Ket{0}$. One defines the Fock vacuum as the vector $\Ket{0}=\Ket{0,0,0,...,0}$ corresponding to $N$ unoccupied states, which is defined by
\begin{equation}
\label{eq:defb}
b_i\,\Ket{0}\,=\,0\,, \text{for all } i=1,...,N\,.
\end{equation}
One-state vector can then be represented as
\begin{equation}
b^{\dagger}_i\,\Ket{0}\,=\, \Ket{0,0,...,1,...,0}\,,
\end{equation}
where the non-zero entry is at position $i$. We have just derived the building blocks to construct the spinor representation of $\mathsf{SO}(2N)$ in terms of states obtained from the action of the 
creation operators. Moreover, defining the set of operators $T_{ij}\equiv b^{\dagger}_i\,b_j$, it is easy to verify that they satisfy the algebra of $\mathsf{U}(N)$ as
\begin{equation}
\left[T_{ij},\,T_{kl}\right]\,=\,\delta_{kj}T_{il}\,-\,\delta_{il}T_{kj}\,.
\end{equation}
It is then not surprising to observe that the basis of vectors obtained through the action of products of creation operators on the Fock vacuum, 
\begin{equation}
\label{eq:vectorbasis}
b^{\dagger}_{i_1}\, b^{\dagger}_{i_2}\,...\, b^{\dagger}_{i_p}\Ket{0}\,,\quad i_1<i_2<\dots<i_p,
\end{equation}
expands any vector $\Ket{\Psi}$ with coefficients being irreducible fully-antisymmetric $\mathsf{U}(N)$ tensors, $\psi^{i_1...i_p}$. This fact allows us to write the spinor representations of $\mathsf{SO}(2N)$ in terms of irreducible $\mathsf{SU}(N)$ tensors.

\subsection{Decomposition into the $\mathsf{SU}(N)$ basis}
\label{sec:dec}

The general expression for the spinor representation of $\mathsf{SO}(2N)$ written in terms of the $\mathsf{SU}(N)$ fields is given by
\begin{equation}
\label{eq:generalpsi}
\Ket{\Psi}\,=\,\sum^N_{p=0}\,\frac{1}{p!}\,b^{\dagger}_{i_1}\, ...\, b^{\dagger}_{i_p}\Ket{0}\,\psi^{i_1...i_p}\,.
\end{equation}
The completely antisymmetric tensors $\psi^{i_1...i_p}$ have dimension $\binom{N}{p}$. An easy way to compute the dimension of all tensors in \cref{eq:generalpsi} is by noting that it can be read from the $N$th-row of the Tartaglia's triangle\footnote{This mathematical representation is also known as the Pascal's triangle. The triangle was already known centuries before in China, India and Iran.}. For tensors with large number of indices it may be convenient to reduce them with help of their conjugate tensors using the Levi-Civita invariant tensor of dimension $N$,
\begin{equation}
\overline{\psi}_{i_{p+1}\dots i_N}=\frac{1}{p!}\,\varepsilon_{i_1\dots i_N}\,\psi^{i_1...i_p}\,,
\end{equation}
and therefore \cref{eq:generalpsi} becomes
\begin{equation}
\label{eq:generalpsi2}
\begin{split}
\Ket{\Psi} &\,=\, \Ket{0}\psi\,+\,b^{\dagger}_i\Ket{0}\psi^i\,+\, \frac{1}{2}\, b^{\dagger}_i b^{\dagger}_j\Ket{0}\psi^{ij}\\
&+\,\cdots\\[2mm]
&+\, \frac{\varepsilon^{i_1i_2\cdots i_N}}{2!(N-2)!}\, b^{\dagger}_{i_1} b^{\dagger}_{i_2}\cdots b^{\dagger}_{i_{N-2}} \Ket{0}\bar{\psi}_{i_{N-1}i_N}\\[2mm]
&+\, \frac{\varepsilon^{i_1i_2\cdots i_N}}{(N-1)!}\, b^{\dagger}_{i_1} b^{\dagger}_{i_2}\cdots b^{\dagger}_{i_{N-1}}\Ket{0}\bar{\psi}
_{i_N}\\[2mm]
&+\, b^{\dagger}_{1} b^{\dagger}_{2}\cdots b^{\dagger}_{N} \Ket{0}\bar{\psi}\,.
\end{split}
\end{equation}
The dimension of the vector space in \cref{eq:generalpsi2} is $2^N$ which is in agreement with the dimension of the $\Gamma$-matrices. Within the $\mathsf{SU}(N)$ basis, given by the vectors of~\cref{eq:vectorbasis}, any spinor $\Ket{\Psi}$ corresponds to a column vector $\Psi$,
\begin{equation}
\Psi=\begin{pmatrix}
\psi&
\psi^{j}&
\psi^{jk}&
\cdots&
\overline{\psi}_{jk}&
\overline{\psi}_{j}&
\overline{\psi}
\end{pmatrix}^{\intercal}\,.
\end{equation}
In this spinor representation of dimension $2^N$, the states $\Psi$ transform under $\mathsf{SO}(2N)$ as 
\begin{equation}
\Psi\,\longrightarrow\,\Psi'=U(\omega)\,\Psi\quad \text{or}\quad
\Ket{\Psi'}=U(\omega)\Ket{\Psi}\,,
\end{equation}
where the unitary transformation $U(\omega)$ is given by
\begin{equation}
U(\omega)=\exp\left(-\frac{i}{2}\omega_{\mu \nu}\Sigma_{\mu \nu}\right)\,.
\end{equation}
The generators $\Sigma_{\mu \nu}$ of the spinor representation are constructed in terms of the $\Gamma$-matrices as 
\begin{equation}
\label{eq:sigma}
\Sigma_{\mu \nu}=\frac{1}{2i}\left[\Gamma_{\mu},\,\Gamma_{\nu}\right]\,,
\end{equation}
with $\Sigma^{\dagger}_{\mu \nu}=\Sigma_{\mu \nu}$ and $\tr\Sigma_{\mu \nu}=0$, which guaranties the unitarity of $U(\omega)$ and $|U(\omega)|=1$, respectively. It is straightforward to verify that $\Sigma_{\mu \nu}$ satisfies the algebra of $\mathsf{SO}(2N)$ given in \cref{eq:algebra}. It turns out that the spinor representation $\Ket{\Psi}$ with dimension $2^{N}$ given in \cref{eq:generalpsi2} is in fact reducible. This fact can easily be demonstrate by observing that the product of $2N$ $\Gamma$-matrices, $\Gamma_0$, defined as
\begin{equation}
\label{eq:gamma0}
\Gamma_0\,=\,i^N \Gamma_1\,\Gamma_2\, ...\, \Gamma_{2N}\,,
\end{equation}
anticommutes with all $\Gamma_{\mu}$ matrices, but it commutes with $\Sigma_{\mu\nu}$ and therefore splits the spinor $\Psi$ into two nonequivalent irreducible spinors $\Psi_{+}$ and $\Psi_{-}$ of dimension $2^{N/2}$, given by
\begin{equation}
\label{eq:projectors}
\Psi_{\pm}=\frac{1}{2}(1\pm \Gamma_0)\Psi\,.
\end{equation}
The projectors $\frac{1}{2}(1\pm \Gamma_0)$ are in total analogy with the chiral projectors known in the Dirac space (see projector properties in \cref{sec:clifford}). The chiral states $\Psi_{+}$ are generated by the action of an even number of creation operators on the vacuum state $\Ket{0}$, while the chiral states $\Psi_{-}$ are generated by the action of an odd number of creation operators. Observing \cref{eq:generalpsi2} one obtains
\begin{equation}
\Psi_{+}=\begin{pmatrix}
\psi\\
\psi^{ij}\\
\overline{\psi}_{i}\\
\vdots
\end{pmatrix}\,,\quad \Psi_{-}=\begin{pmatrix}
\overline{\psi}\\
\overline{\psi}_{ij}\\
\psi^{i}\\
\vdots
\end{pmatrix}\,,
\end{equation}
and one can distinguish two cases in $\mathsf{SO}(2N)$ : when $N$ is an odd integer $\Psi_{+}$ and $\Psi_{-}$ are self-conjugate, while when $N$ is even $\Psi_{+}$ and $\Psi_{-}$ are distinct spinor representations.

%%%%%%%%%%%%%%%%%%%%%%%%%%%%%%%%%%%%%%
\subsection{Yukawa interactions}
\label{subsec:Yukawa}

In this subsection, we sketch the construction of Yukawa interactions in a generic GUT model ruled by $\mathsf{SO}(2N)$ expressed in terms of $\mathsf{SU}(2N)$ tensor fields. This is particularly useful when the group $\mathsf{SO}(2N)$ is broken to its $\mathsf{SU}(2N)$ subgroup at some intermediate scale, since below the breaking scale one gets new relations among the Yukawa couplings of the $\mathsf{SU}(N)$ theory. In what follows, we shall assume that the fermionic degrees of freedom belong to irreducible $\mathsf{SO}(2N)$ spinor representations and the Higgs fields transform as complete antisymmetric tensors of $\mathsf{SO}(2N)$. Since the GUT gauge group commutes with space-time symmetries, it is convenient to write the fermionic degrees of freedom in terms of left-handed Weyl fields. In order to write the most general Yukawa interaction invariant under $\mathsf{SO}(2N)$, it is useful to express the transposition of $U(\omega)$ in terms of its corresponding inverse matrix $U^{\dagger}(\omega)$ as
\begin{equation}
\label{eq:needB}
U^{\intercal}(\omega)=B\,U^{\dagger}(\omega)\,B^{-1}\,,
\end{equation}
through the matrix $B$, which has the property
\begin{equation}
 B^{-1}\,\Gamma_{\mu}^{\intercal}\,B\,=\,\Gamma_{\mu}\,.
\end{equation}
In \cref{sec:clifford}, it is shown that such operator $B$ always exists and it can be written as 
\begin{equation}
\label{eq:Bopgen}
B=\prod_{\mu=odd} \Gamma_{\mu}=(-i)^N\prod_{k=1}^{N}(b_k-b^{\dagger}_k)\,.
\end{equation}
From the above equation, one sees that the operator $B$ anticommutes with $\Gamma_0$ when $N$ is odd, while instead it commutes when $N$ is even. When simplifying expressions involving the operator $B$, it turns out to be more convenient to write it with contracted indices as
\begin{equation}
\label{eq:B}
B=\frac{(-i)^N}{N!}\,\varepsilon^{jk\cdots z}\,(b_j-b^{\dagger}_j)(b_k-b^{\dagger}_k)\cdots (b_z-b^{\dagger}_z)\,,
\end{equation}
where $\varepsilon^{jk\cdots z}$ is the Levi-Civita antisymmetric tensor with $N$ indices. From \cref{eq:needB}, one deduces that the combination $\Psi^{\intercal}B$ does transform as 
\begin{equation}
\Psi^{\intercal}B\,\longrightarrow\,{\Psi'}^{\intercal}B=\Psi^{\intercal}B\,U^{\dagger}(\omega)\,,
\end{equation}
and one concludes that for any pair of spinors $\Psi_1$ and $\Psi_2$ the bilinear $\Psi_1^{\intercal}BC^{-1}\,\Psi_2$ is invariant under $\mathsf{SO}(2N)$. Due to the fact that $\Psi_1$ and $\Psi_2$ are assumed as fermionic fields, the presence of charge conjugation matrix $C$ ensures that this bilinear is also invariant under Lorentz transformations. Furthermore, the matrices $\Gamma_{\mu}$ transform as
\begin{equation}
\label{eq:gammatrans}
U(\omega)^{\dagger}\,\Gamma_{\mu}\, U(\omega) \,=\, O_{\mu\nu}\, \Gamma_{\nu}\,,
\end{equation}
where the matrix $O$ is given in \cref{eq:O}. Using this result one can write an $\mathsf{SO}(2N)$ invariant Yukawa coupling combining the fermion in a spinor representation $\Psi$ with the Higgs scalar $\phi_{\mu}$, that transforms like a vector according to \cref{eq:genSO}:
\begin{equation}
\Psi^{\intercal}B\,C^{-1}\Gamma_{\mu}\,\Psi\;
\phi_{\mu}\,.
\end{equation}

The relation given in \cref{eq:gammatrans} can be generalised to any product of $\Gamma$-matrices as
\begin{equation}
\begin{aligned}
U(\omega)^{\dagger}\,\Gamma_{\mu}\Gamma_{\nu}\cdots\Gamma_{\rho}\, U(\omega) =
O_{\mu\mu'} O_{\nu\nu'}\cdots O_{\rho\rho'} \Gamma_{\mu'}\Gamma_{\nu'}\cdots\Gamma_{\rho'}\,.
\end{aligned}
\end{equation}
This general formula allows us to write the most general gauge-invariant Yukawa coupling under $\mathsf{SO}(2N)$. In the language of creation and annihilation operators one writes
\begin{equation}
\label{eq:yukawaGen}
Y_{ab}\,\Bra{\Psi^{\ast}_{k\, a}}B\,C^{-1}\,\Gamma_{[\mu_1}\Gamma_{\mu_2}\,{}_{\cdots}\,\Gamma_{\mu_m]}\,\Ket{\Psi_{l\, b}}\, \phi_{{\mu_1}{\mu_2}\cdots{\mu_m}}\,,
\end{equation}
where the indices $k,l=+,-$ denote the two possible irreducible spinors; $a$ and $b$ are flavor indices and the elements $Y_{ab}$ of the Yukawa matrix; $\phi_{{\mu_1}{\mu_2}\cdots{\mu_m}}$ with $m=1,\dots N$ is a scalar tensor fully antisymmetric. Due to the fact that $\phi_{{\mu_1}{\mu_2}\cdots{\mu_m}}$ is fully antisymmetric, any $\Gamma$-matrix product in the formula in \cref{eq:yukawaGen} should also be made fully antisymmetric, and therefore one has
\begin{equation}
\begin{aligned}
\Gamma_{[\mu_1}\Gamma_{\mu_2}{}_{\cdots}\Gamma_{\mu_m]}\,=\,\frac{1}{m!}
\sum_{P}(-1)^{\delta P} \Gamma_{\mu_{P(1)}}\Gamma_{\mu_{P(2)}}\cdots\Gamma_{\mu_{P(m)}}\,,
\end{aligned}
\end{equation}
where the sum runs over the permutations and $\delta P$ takes 0 for even number of permutations and 1 for odd number of permutations. We notice that the general formula given in \cref{eq:yukawaGen} can also be applied to the case where the $\mathsf{SO}(2N)$ spinors are taken as scalar fields yielding a pure scalar interaction in the scalar potential. In this case, the bracket in \cref{eq:yukawaGen} should not include the charge conjugation matrix $C$.

In some cases, the bracket given in \cref{eq:yukawaGen} vanishes automatically. This can be well understood by taking into account the general properties of the projectors $\frac{1}{2}(1\pm \Gamma_0)$ and the fact that $\Gamma_0$ commutes with an even number of $\Gamma$-matrix product or anticommutes with an odd number. Thus, in the case $N+m$ is an odd number, it implies 
\begin{equation}
\Bra{\Psi^{\ast}_{k\,a}}B\,C^{-1}\,\Gamma_{\mu_1}\Gamma_{\mu_2}\cdots\Gamma_{\mu_m}\,\Ket{\Psi_{k\,a}}\,=\,0\,,
\end{equation}
while in the case $N+m$ is an even number, it implies 
\begin{equation}
\Bra{\Psi^{\ast}_{k\,a}}B\,C^{-1}\,\Gamma_{\mu_1}\Gamma_{\mu_2}\cdots\Gamma_{\mu_m}\,\Ket{\Psi_{l\,b}}\,=\,0\,,
\end{equation}
when $k\neq l$.

Concerning the antisymmetric tensor $\phi_{{\mu_1}{\mu_2}\cdots{\mu_m}}$ some comments are in order. There are $N$ distinct fully antisymmetric tensors with dimension $\binom{2N}{m}$ with $1\leq m\leq N$\footnote{Representations of dimension $\binom{2N}{m}$ with $m>N$ are equivalent to representations with dimension $\binom{2N}{m-N}$.}. Moreover, the representation with dimension $\binom{2N}{N}$, denoted as $\Delta_ {\mu_1\,\mu_2\,\cdots\,\mu_N}$, is indeed a reducible representation that can be decomposed into two irreducible representations~\cite{Harvey:1980je,Harvey:1981hk}. For $N$ odd into two irreducible pairs self-conjugate representations of dimension $\tfrac12\binom{2N}{N}$:
\begin{equation}
\Delta_{\mu_1\,\mu_2\,\cdots\,\mu_N}=\overline{\phi}_{\mu_1\,\mu_2\,\cdots\,\mu_N}\,+\,
\phi_{\mu_1\,\mu_2\,\cdots\,\mu_N}\,,
\end{equation}
where 
\begin{equation}
\label{eq:256red}
\begin{pmatrix}
\overline{\phi}_{\mu_1\cdots\mu_N}\\
\phi_{\mu_1\cdots\mu_N}
\end{pmatrix} \equiv\frac12\,
\begin{pmatrix}
\delta_{\mu_1\nu_1}\cdots\delta_{\mu_N\nu_N} + \frac{i}{N!}\varepsilon_{\mu_1\cdots\mu_N\nu_1\cdots\nu_N} \\
\delta_{\mu_1\nu_1}\cdots\delta_{\mu_N\nu_N} - \frac{i}{N!}\varepsilon_{\mu_1\cdots\mu_N\nu_1\cdots\nu_N}
\end{pmatrix}
\,\Delta_{\nu_1\cdots\nu_N}\,.
\end{equation}
Instead, for $N$ even, one has the following decomposition
\begin{equation}
\Delta_{\mu_1\,\mu_2\,\cdots\,\mu_N}={\phi^{+}}_{\mu_1\,\mu_2\,\cdots\,\mu_N}\,+\,
\phi^{-}_{\mu_1\,\mu_2\,\cdots\,\mu_N}\,,
\end{equation}
where
\begin{equation}
\label{eq:256red}
\begin{pmatrix}
{\phi}^{+}_{\mu_1\cdots\mu_N}\\
\phi^{-}_{\mu_1\cdots\mu_N}
\end{pmatrix} \equiv\frac12\,
\begin{pmatrix}
\delta_{\mu_1\nu_1}\cdots\delta_{\mu_N\nu_N} + \frac{1}{N!}\varepsilon_{\mu_1\cdots\mu_N\nu_1\cdots\nu_N} \\
\delta_{\mu_1\nu_1}\cdots\delta_{\mu_N\nu_N} - \frac{1}{N!}\varepsilon_{\mu_1\cdots\mu_N\nu_1\cdots\nu_N}
\end{pmatrix}
\,\Delta_{\nu_1\cdots\nu_N}\,.
\end{equation}
When computing the full expression given by~\cref{eq:yukawaGen} for the maximal number of $\Gamma$-matrices, one verifies that only one of the irreducible components of $\Delta_{\mu_1\,\cdots\,\mu_N}$ couples to the Yukawa term. Indeed, for $N$ odd, one has $k=l$ and therefore only $\overline{\phi}_{\mu_1\cdots\mu_N}$($\phi_{\mu_1\cdots\mu_N}$) couples to the Yukawa when $k=+$($-$), otherwise, for $N$ even, one has $k\neq l$ and therefore only $\phi^{+}_{\mu_1\cdots\mu_N}$($\phi^{-}_{\mu_1\cdots\mu_N}$) couples to the Yukawa when $k=-$($+$). One then concludes that the computation of the Yukawa given by~\cref{eq:yukawaGen} can be performed directly using the reducible representation $\Delta_{\mu_1\,\mu_2\,\cdots\,\mu_N}$ as
\begin{equation}
Y_{ab}\,\Bra{\Psi^{\ast}_{k\, a}}B\,C^{-1}\,\Gamma_{[\mu_1}\Gamma_{\mu_2}\,{}_{\cdots}\,\Gamma_{\mu_N]}\,\Ket{\Psi_{l\, b}}\, \Delta_{\mu_1\,\cdots\,\mu_N}\,,
\end{equation}
without the loss of generality.

For illustrative purpose, the antisymmetric tensor representations have the following dimension in the case of $\mathsf{SO}(10)$: $\phi_{\mu}\sim\mathsf{10}$, $\phi_{\mu\nu}\sim\mathsf{45}$, $\phi_{\mu\nu\lambda}\sim\mathsf{120}$, $\phi_{\mu\nu\lambda\sigma}\sim\mathsf{210}$, $\phi_{\mu\nu\lambda\sigma\gamma}\sim\mathsf{126}$, and $\overline{\phi}_{\mu\nu\lambda\sigma\gamma}\sim\overline{\mathsf{126}}$.

One can also express the tensor $\phi_{\mu \nu \lambda...\sigma}$ in terms of $\mathsf{SU}(N)$ tensors. This can be easily computed by expanding the quantity $\Gamma_{\mu}\Gamma_{\nu}\Gamma_{\lambda}...\Gamma_{\sigma}\,\phi_{\mu \nu \lambda...\sigma}$ in terms of the creation and annihilation operators~\cite{Nath:2001uw} as
\begin{equation}
\label{eq:basictheo}
\begin{aligned}
\Gamma_{\mu}\Gamma_{\nu}&\Gamma_{\lambda}...\Gamma_{\sigma}
\phi_{\mu \nu \lambda...\sigma}=b^{\dagger}_ib^{\dagger}_jb^{\dagger}_k...b^{\dagger}_n\phi_{c_ic_jc_k...c_n}\\
&+(b_ib^{\dagger}_jb^{\dagger}_k...b^{\dagger}_n\phi_{\bar{c}_ic_jc_k...c_n} + \text{perms})\\
&+(b_ib_jb^{\dagger}_k...b^{\dagger}_n\phi_{\bar{c}_i\bar{c}_jc_k...c_n} + \text{perms})+ ... \\
&+(b_ib_jb_k...b_{n-1}b^{\dagger}_n\phi_{\bar{c}_i\bar{c}_j\bar{c}_k...\bar{c}_{n-1}c_n} + \text{perms}) \\
&+(b_ib_jb_k...b_n\phi_{\bar{c}_i\bar{c}_j\bar{c}_k...\bar{c}_n})\,,
\end{aligned}
\end{equation}
where $\phi_{\cdots c_j\cdots}\equiv\phi_{\cdots2j\cdots}+i\,\phi_{\cdots2j-1\cdots}$ and $\phi_{\cdots\bar{c}_j\cdots}\equiv\phi_{\cdots2j\cdots}-i\,\phi_{\cdots2j-1\cdots}$. The new tensors in equation \cref{eq:basictheo} manifest completely antisymmetry, i.e.,
\begin{equation}
\begin{aligned}
\phi_{\cdots c_i\cdots c_j\cdots}&=-\phi_{\cdots c_j\cdots c_i\cdots}\,,\\
\phi_{\cdots \bar{c}_i\cdots \bar{c}_j\cdots}&=-\phi_{\cdots \bar{c}_j\cdots \bar{c}_i\cdots}\,,\\
\phi_{\cdots c_i\cdots \bar{c}_j\cdots}&=-\phi_{\cdots \bar{c}_j\cdots c_i\cdots}\,.
\end{aligned}
\end{equation}
 In \cref{sec:appSO10}, we compile all the antisymmetric tensors of $\mathsf{SO}(10)$ explicitly written in terms of $\mathsf{SU}(5)$ representations.

%%%%%%%%%%%%%%%%%%%%%%%%%%%%%%%%%%%%%%
\subsection{Methods and rules}
\label{subsec:comp} 

We compile in this section the rules that are the basis for defining the behaviour of the \SOSpin library. In order to concretise the use of the rules, let us first take a simple bracket containing a set of annihilation and creator operators, 
\begin{equation}
\label{eq:example}
\Upsilon^{ijl}_{kmn}\,=\,\Bra{0}\,b_i\,b_j\,{b^\dagger_k}\,b_l\,{b^\dagger_m}\, {b^\dagger_n}\Ket{0}\,.
\end{equation}
The tensor $\Upsilon^{ijl}_{kmn}$ differentiates upper and lower indices associated with annihilation and creation operator indices, respectively. This distinction is important to obtain a final expression with a index structure consistent with $\mathsf{SU}(N)$. The computation of the bracket from \cref{eq:example} relies on the use of the relation given in \cref{eq:anticomrelation,eq:defb}. We present two different strategies, that we call \emph{normal ordering} and \emph{reverse ordering} methods. We first discuss the reverse ordering method.

\subsubsection*{Reverse ordering method}
In this case, we move according to \cref{eq:anticomrelation} either the annihilation operators forward to the right until it cancels with $\Ket{0}$, $b_i\Ket{0}$=0, or the creation operators backward to the left to cancel with $\Bra{0}$, i.e., $\Bra{0}b^\dagger_i=0$. 
Hence, moving the operator $b_l$ to the right, we have
\begin{equation}
\begin{aligned}
&\Bra{0}\,b_i\,b_j\,
{b^\dagger_k}\,b_l\,{b^\dagger_m}\, {b^\dagger_n}\Ket{0}=\\
=&\Bra{0}\,b_i\,b_j\,b^\dagger_k\,
\left({\delta_{lm}}-{b^\dagger_m}b_l\right)\, {b^\dagger_n}\Ket{0}\\
&...\\
=&\,\delta_{lm} \Bra{0}\,b_i\,b_j\,{b^\dagger_k}\,{b^\dagger_n}\Ket{0}-{\delta_{ln}} \Bra{0}\,b_i\,b_j\,{b^\dagger_k}\,{b^\dagger_m}
\Ket{0}\,.\\
\end{aligned}
\end{equation}

We complete the computation of $\Bra{0}\,b_i\,b_j\,{b^\dagger_k}\,{b^\dagger_n}\Ket{0}$ and $\Bra{0}\,b_i\,b_j\,{b^\dagger_k}\,{b^\dagger_m}
\Ket{0}$ using the same procedure, obtaining 
\begin{equation}
\begin{aligned}
&\Bra{0}\,b_i\,b_j\,{b^\dagger_k}\,{b^\dagger_n}\Ket{0}
=\,\delta_{jk}\,\delta_{in}-
\delta_{jn}\,\delta_{ik}\,,\\
&\Bra{0}\,b_i\,b_j\,{b^\dagger_k}\,{b^\dagger_m}\Ket{0}
=\,\delta_{jk}\,\delta_{im}-\delta_{jm}\,\delta_{ik}\,.
\end{aligned}
\end{equation}

The final result for $\Bra{0}\,b_i\,b_j\,
{b^\dagger_k}\,b_l\,{b^\dagger_m}\, {b^\dagger_n}\Ket{0}
$, in terms of $\delta$'s, is then given by
\begin{equation}
\delta_{lm} (\delta_{jk}\,\delta_{in}-
\delta_{jn}\,\delta_{ik}) -\delta_{ln}(\delta_{jk}\,\delta_{im}-
\delta_{jm}\,\delta_{ik})\,.
\end{equation}

As already said, instead we move the annihilation operators to the right-handed side to cancel at $\Ket{0}$, we can move the creation operators to the left to cancel when reach $\Bra{0}$. However, once we choose to move either annihilation or creation operators, we need to maintain this choice until the end of the computation.

\subsubsection*{Normal ordering method}
In the normal ordering method, we use the relations in \cref{eq:anticomrelation} to rearrange the creation and annihilation operators in such a way that all annihilation operators are on the left-handed side while all creation operators are on the right-handed side, as
\begin{equation}
\label{eq:semi}
\begin{aligned}
&\Bra{0}\,b_i\,b_j\,b^\dagger_k\,b_l\,{b^\dagger_m}\, {b^\dagger_n}\Ket{0}=\\
=&\Bra{0}\,b_i\,b_j\,(\delta_{kl}-\,b_l\,{b^\dagger_k})\,{b^\dagger_m}\, {b^\dagger_n}\Ket{0}\\
=&\,{\delta_{kl}}\Bra{0}\,b_i\,b_j\,{b^\dagger_m}\, {b^\dagger_n}\Ket{0}-\Bra{0}\,b_i\,b_j\,\,b_l\,{b^\dagger_k}\,{b^\dagger_m}\, {b^\dagger_n}\Ket{0}\,.
\end{aligned}
\end{equation} 
Then we compute $\Bra{0}\,b_i\,b_j\,b_l\,{b^\dagger_k}\,{b^\dagger_m}\, {b^\dagger_n}\Ket{0}$ and $\Bra{0}\,b_i\,b_j\,{b^\dagger_m}\, {b^\dagger_n}\Ket{0}$ by using the relation,
\begin{equation}
\label{eq:general}
\begin{aligned}
&\Bra{0}b_{i_1}b_{i_2}...b_{i_k}\, b^\dagger_{j_1}b^\dagger_{j_2}...b^\dagger_{j_k}\Ket{0}=\\ 
&\frac{1}{(N-k)!}\, \varepsilon^{{i_1}{i_2}...{i_k}{l_{k+1}}...{l_N}}\, \varepsilon_{{j_k}...{j_2}{j_1}{l_{k+1}}...{l_N}}\,,
\end{aligned}
\end{equation}
that holds for $\mathsf{SU}(N)$ with $k\leq N$.

Up to now, we did not mention in which $\mathsf{SU}(N)$ framework we are computing the expression given in \cref{eq:example}. If we choose to compute it in $\mathsf{SU}(5)$ (i.e., $N=5$) the brackets in \Cref{eq:semi} take the values
\begin{equation}
\begin{aligned}
&\Bra{0}\,b_i\,b_j\,{b^\dagger_m}\,{b^\dagger_n}\Ket{0}=\frac{1}{3!}\, \varepsilon^{ij\alpha\beta\gamma}\, \varepsilon_{nm\alpha\beta\gamma}\,,\\
&\Bra{0}\,b_i\,b_j\,b_l\,b^\dagger_k\,{b^\dagger_m}\,{b^\dagger_n}\Ket{0}=\frac{1}{2!}\,\varepsilon^{ijl\alpha\beta}\, \varepsilon_{nmk\alpha\beta}\,,
\end{aligned}
\end{equation}
and the tensor $\Upsilon^{ijl}_{kmn}$ defined in \cref{eq:example} becomes then
\begin{equation}
\Upsilon^{ijl}_{kmn}\,=\,\frac{1}{6} \,\delta^{l}_{k}\, \varepsilon^{ij\alpha\beta\gamma}\, \varepsilon_{nm\alpha\beta\gamma}\,-\, \frac{1}{2} \varepsilon^{ijl\alpha\beta}\, \varepsilon_{nmk\alpha\beta}.
\end{equation} 

We are ready to summarise all the rules mentioned above. It is known that a complete bracket expression is composed by creation and annihilation operators, and some other fields. The idea is to get rid of all operators by using the relations in \cref{eq:anticomrelation} and use the terms arising from these calculations to simplify the remaining expression (e.g., fields if there are any). In order to have a consistent expression, each element in the bracket must obey certain rules, those that we list below for a general $\mathsf{SO}(2N)$:
\begin{enumerate}
	\item The number of creation operators in a bracket expression must be equal to the number of annihilation operators.
	\item The number of contiguous creation (or annihilation) operators must be equal or less than $N$ for $\mathsf{SO}(2N)$. 
	\item The operators $b_i$ inside the bracket expression are written on the left-handed side while the operators $b^\dagger_i$ are written on the right-handed side, otherwise the result is zero, i.e., $b_i\ket{0}=0$ and $\bra{0}b^\dagger_i=0$.
	\item The difference between the number of upper and lower indices in the fields must be zero or multiple of $N$ for any $\mathsf{SO}(2N)$.
\end{enumerate}

Within this framework, we shall give the \emph{Hermitian conjugation} and the \emph{transposition} operations on a general vector
\begin{equation}
\begin{aligned}
\Ket{\Psi}\,&=\,\Ket{0}\,\psi 
\,+\, 
b^{\dagger}_i\Ket{0} \,\psi^i
\,+\,\frac{1}{2}b^{\dagger}_ib^{\dagger}_j\Ket{0}\,\psi^{ij}\\
&\,+\,\frac{1}{12}\varepsilon^{ijklm}b^{\dagger}_kb^{\dagger}_lb^{\dagger}_m\Ket{0}\,\psi_{ij}
\,+\, ... \,,
\end{aligned}
\end{equation}
which are
\begin{equation}
\begin{aligned}
\Bra{\Psi}\,&=\,{\psi}^\ast\Bra{0} 
\,+\, 
\,{\psi_i} \Bra{0} b_i
\,+\,\frac{1}{2}{\psi_{ij}}\, \Bra{0}\, b_jb_i\\
&\,+\,\frac{1}{12}\,\psi^{ij}\varepsilon_{ijklm}\Bra{0}b_mb_lb_k\,+\, ...\,,
\end{aligned}
\end{equation}
and 
\begin{equation}
\begin{aligned}
\Bra{\Psi^{\ast}}\,%&
=\,\psi\Bra{0} 
\,+\, 
\,{\psi^i} \Bra{0} b_i
\,+\,\frac{1}{2}{\psi^{ij}}\, \Bra{0}\, b_jb_i%\\
%&
\,+\,\frac{1}{12}\psi_{ij}\varepsilon^{ijklm}\Bra{0}\,b_m\,b_lb_k	
\,+\, ...\,,
\end{aligned}
\end{equation}
respectively, where we make the usual identification of lower indices as $\psi_{ij}=\left(\psi^{ij}\right)^\ast$.

%%%%%%%%%%%%%%%%%%%%%%%%%%%%%%%%%%%%%%
\section{ \texorpdfstring{\protect\SOSpin}{SOSpin}, a \Cpp library}
\label{sec:prog} 

\begin{figure}[t]
	\centering
	\includegraphics[width=0.8\linewidth]{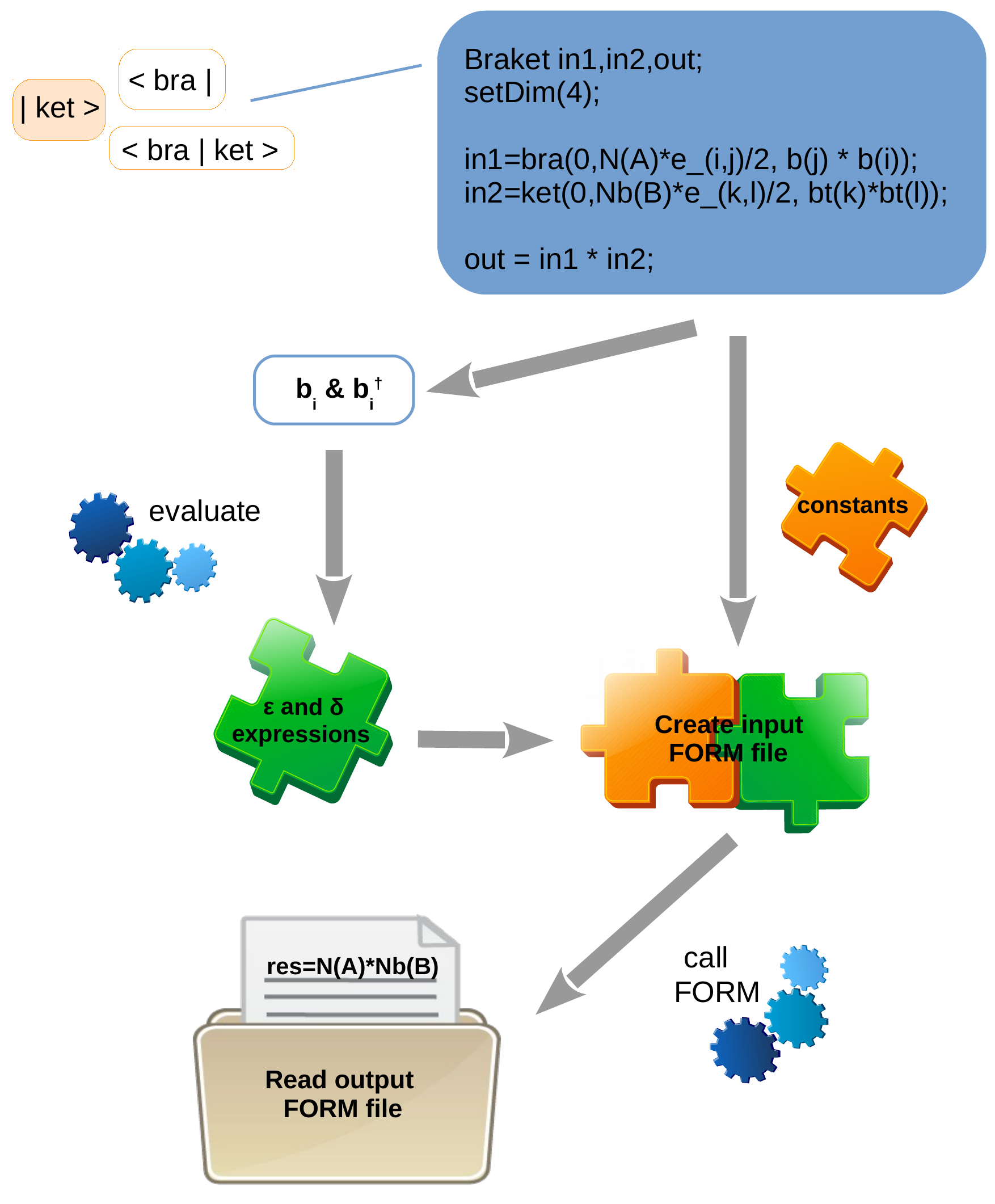}
	\caption{\label{fig:scheme} Scheme of \protect\SOSpin library exemplifying how to compute $\frac{\varepsilon_{ij}\varepsilon_{kl}}{4}N_AN_B\bra{0} b_jb_i b^\dagger_k b^{\dagger}_l \ket{0}$ in $\mathsf{S0}(4)$.}
\end{figure}

In this section we present the structure of the \SOSpin library, that can be found in \url{http://sospin.hepforge.org}, comment on the data structure representation and give a list of the most important functions to use when writing a program linked to the \SOSpin library. As mentioned before, the \SOSpin code is entirely written in \Cpp and it is based on operations over creation and annihilation operators. We give in \cref{fig:scheme} the pictorial scheme to explain how a \SOSpin program works. It works like this: the building blocks of the code, i.e., the $\bra{bra}$ and the $\ket{ket}$ entities, can be defined either in a main file program (\code{in1} and \code{in2} in the picture) or included in some \code{include} file and called in the main file. The implemented operations in the \SOSpin library are the following:
\begin{itemize}
	\item $\bra{bra} \cdot \ket{ket}$
	\item $\bra{bra} \cdot$ free
	\item free $\cdot \ket{ket}$
	\item $\bra{bra} + \bra{bra}$
	\item $\ket{ket} + \ket{ket}$
	\item free + free
	\item free $\cdot$ free
	\item $\braket{bra|ket} + \braket{bra|ket}$
	\item $\braket{bra|ket} \cdot \braket{bra|ket}$
	[This operation is only allowed after the evaluation of the expression.]
\end{itemize}

Once the expression to evaluate is defined (\code{out} in the picture) the approach to solve it is the following: the expression to evaluate is split in two parts, one with all constants and another one containing operators; the operator part will be worked out using one of the two methods described in \cref{sec:dec} leading to an intermediate expression written in terms of $\varepsilon$'s or $\delta$'s. Then, the constant part and the intermediate expression are joined together to lead to a semi-final expression. If the starting example is simple, the result will be simple. On the other hand, if the example is somewhat complex, the expression obtained at this stage is rather large and needs extra simplifications for increasing the readability of the final expressions. Hence, in order to make the reading of all results as easier as possible, we include the possibility to simplify the expression obtained so far with the Symbolic Manipulation System FORM~\cite{Kuipers:2012rf}. Once the constant part and the intermediate expression are joined together, it is created an input file to be run by FORM leading to a more simplified expression. The final expression is then read back from the FORM output file to the program. Note that, in the output FORM file, we first present all partial results and then the complete result at the end of the file.

\begin{figure}[t]
	\centering
	\includegraphics[width=0.8\linewidth]{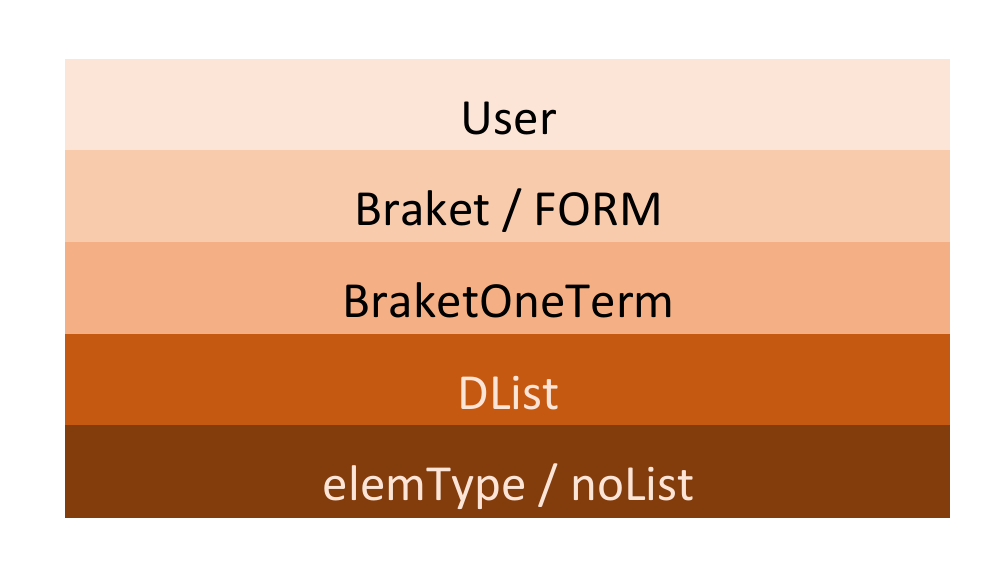}
	\caption{\label{fig:scheme1} Hierarchy of the \protect\SOSpin library.}
\end{figure}

We have opted to include the FORM program as a tool to do the final simplifications if needed. For the sake of curiosity we have implemented in FORM all the procedure used in \SOSpin; however, the running time measured is far larger than in the case of using our library in \Cpp. This shows the performance power of our choice for the appropriate data structure, which is described in the next section.

In \cref{fig:scheme1} we define how the class structure works. In terms of level abstraction the low level implementation of the basic elements for the expression evaluation are the structures \code{elemType} and \code{noList}. They represent the expression elements or nodes. The class \code{DList} is then build over this abstraction level and represents a linked connection of nodes. Then \code{BraketOneTerm} and \code{Braket} classes represent complete expressions that can be evaluated and simplified by FORM and by the User at the higher abstraction level. 

%%%%%%%%%%%%%%%%%%%%%%%%%%%%%%%%%%%%%%
\subsection{Data structure representation}
\label{subsec:data}

\begin{figure*}[t]
\centering\includegraphics{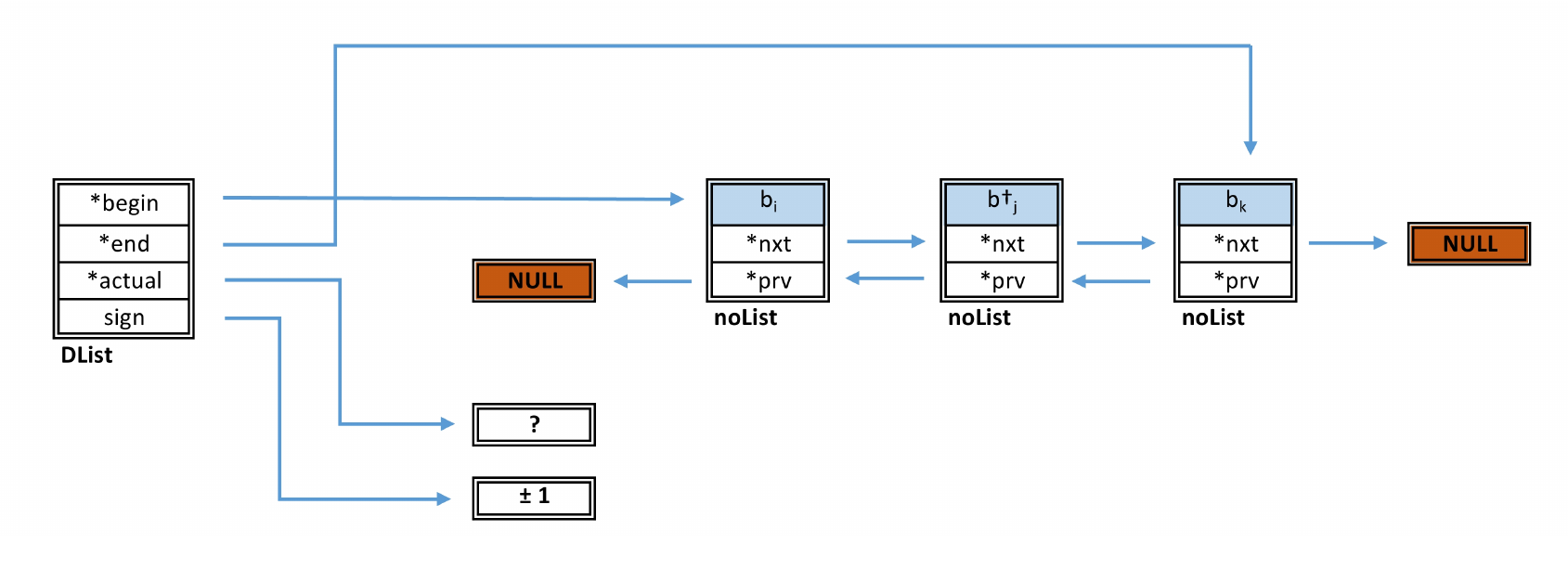}
\caption{\label{fig:DList} Representation of the bracket $\braket{0|b_ib^{\dagger}_jb_k|0}$ in a linked data structure.}
\end{figure*}

In order to manipulate sequences of operators $b_i$ or $b^\dagger_i$, we need to find the adequate data structure to store and further evaluate such sequences. Such data structures should require the following criteria:
\begin{itemize}
\item optimize memory usage - since the sequences can get extremely long;
\item optimize flexibility of permutations - adjacency in memory is not relevant;
\item standardize description of all elements - to ease interpretation and evaluation, and to reduce memory waste in contraction and expansion operations.
\end{itemize}
We have adopted the \textbf{doubly-linked list} scheme as the appropriate solution to the problem. A doubly-linked list consists on the list of connected nodes, which include specific data objects, such as each node is linked to previous and next nodes in the list. This is advantageous since one changes only the pointers without modifying the content and their position on the memory. The doubly-linked list scheme is implemented by a \Cpp class named as \code{DList}. The full method list is included in the~\cref{subsec:dlist}. 
 
The different types of elements within a sequence are encoded as bit-fields of an integer type (\code{int}) with the purpose of optimising the memory usage. Thus, each \code{DList}-node can account for the operators $b_i$ and $b^{\dagger}_i$, as well as the constants and the Kronecker symbol $\delta$, with its indices. In \cref{fig:DList}, we illustrate the concept of the \code{DList} class for a simple sequence.

\paragraph{Computation Performance}
In order to give an estimate of the performance of the computation, we measured the time consumed and the memory used in computing a sequence of creation and annihilation operators. We run the test in a x64 LINUX machine (Ubuntu) with an Intel(R) Core(TM) i5-3317U CPU @ 1.70GHz.

We tested the following expression for $\mathsf{SO}(10)$,
\begin{equation}
\bra{0} b_{i_1} b_{i_2} b^\dagger_{j_1} b_{i_3} b_{i_4} b_{i_5} b^\dagger_{j_2} b^\dagger_{j_3} b_{i_6} b^\dagger_{j_4} b_{i_7} b_{i_8} b^\dagger_{j_5} b^\dagger_{j_6} b^\dagger_{j_7} b^\dagger_{j_8}\ket{0}\,,
\end{equation}
the program needs a total of 3.57 MB and 0.0210 s to evaluate this expression in the delta form and 1.39 MB and 0.0044 s to evaluate the same expression to Levi-Civita tensor form.

Running the following sequence in $\mathsf{SO}(18)$,
\begin{equation}
\bra{0} b_{i_1} b_{i_2} b_{i_3} b_{i_4} b_{i_5} b_{i_6} b_{i_7} b_{i_8} b_{i_9} b^\dagger_{j_1} b^\dagger_{j_2} b^\dagger_{j_3} b^\dagger_{j_4} b^\dagger_{j_5} b^\dagger_{j_6} b^\dagger_{j_7} b^\dagger_{j_8} b^\dagger_{j_9}\ket{0}\,,
\end{equation}
the program needs a total of 454.15 MB and 2.80 s to evaluate this expression in the delta form and 1.41 MB and 0.00064 s to evaluate the same expression to the Levi-Civita tensor form. The large amount of memory used to evaluate in the delta form is due to the number of terms generated in this way, a total of $9!=362880$ terms, while for the evaluation to Levi-Civita tensor the result has only one final term.

%%%%%%%%%%%%%%%%%%%%%%%%%%%%%%%%%%%%%%
\subsection{General functions}
\label{sec:generalfunctions}

In this section we list the most general functions needed to write a program using \SOSpin library; it is divide in three subsections: generic and building functions as well as specific functions to interface with FORM.

%%%%%%%%%%%%%%%%%%%%%%%%%%%%%%%%%%%%%%
\subsubsection*{Generic functions}

\begin{itemize}
\item \code{void setDim(int n)} \\[2mm] 
Sets the group dimension.
\item \code{int getDim()} \\[2mm] 
Gets the group dimension.
\item \code{void CleanGlobalDecl()} \\[2mm] 
Cleans all tables with indices and function declarations.
\item \code{void setVerbosity(Verbosity verb)} \\[2mm] 
Sets verbosity level; verbosity options: SILENT, SUMMARIZE, VERBOSE, DEBUG\_VERBOSE.
\item \code{Verbosity getVerbosity()} \\[2mm] 
Returns current verbosity level.
\end{itemize}

%%%%%%%%%%%%%%%%%%%%%%%%%%%%%%%%%%%%%%
\subsubsection*{Building functions}
\begin{itemize}
	\item \code{DList b(i)}/\code{DList bb(i)}\\[2mm] 
Declares a operator $b_i$; the index in \code{bb()} must be enclosed in quotation marks or passed as a std::string type.
	\item \code{DList bt(i)}/\code{DList bbt(i)}\\[2mm] 
	Declares a operator $b^\dagger_i$; the index in \code{bbt()} must be enclosed in quotation marks or passed as a std::string type.
	\item \code{DList delta(i,j)}\\[2mm] 
	Declares the $\delta_{ij}$ function.
	\item \code{DList identity}\\[2mm] 
	Declares the identity matrix.
	\item \code{Braket bra(A, B, C)}\\
	\code{Braket ket(A, B, C)}\\
	\code{Braket braket(A, B, C)}\\
	\code{Braket free(A, B, C)}\\[2mm] 
	The element A corresponds to the global index, B to the constant part (e.g. fields) and C to the operators $b$ and $b^\dagger$, $\delta$ or the identity. The first entry is the sum of the number of all upper indices (positive counting) and lower indices (negative counting) present in the fields defined in the function. This entry can be set to zero and if so we must call first the function \code{unsetSimplifyIndexSum()}.

\item \code{void evaluate(bool onlydeltas=true)}\\[2mm]
Evaluates expression, if onlydeltas is true then the expression is evaluated to deltas, if false the expression is evaluated to Levi-Civita tensors with eventual $\delta$'s.

\item \code{Braket Bop(std::string startid="i")}\\[2mm] 
Returns the operator $B$ using generic indices.\\[2mm]
\code{Braket BopIdnum()}\\[2mm] 
Returns the operator $B$ using numeric indices.

\item \code{void newId(string i)}\\[2mm] 
Declares a new index.

\item \code{void setSimplifyIndexSum()}\\
\code{void unsetSimplifyIndexSum()}\\[2mm] 
Activates/Deactivates internal simplifications based on the Braket Index sum. This option is activated by default.

\end{itemize}

%%%%%%%%%%%%%%%%%%%%%%%%%%%%%%%%%%%%%%
\subsubsection*{Specific functions to interface with FORM}
\begin{itemize}

\item \code{std::string Field(A, B, C, D)}\\[2mm] 
This function is used to declare the field in FORM, where
\begin{enumerate}
	\item[A] field name;
	\item[B] number of upper indices;
	\item[C] number of lower indices;
	\item[D] field properties:
	\begin{itemize}
		\item SYM: symmetric field without flavor index;
		\item ASYM: antisymmetric field without flavor index;
		\item SYM\_WITH\_FLAVOR: symmetric field with flavor index;
		\item ASYM\_WITH\_FLAVOR: antisymmetric field with flavor index.
	\end{itemize}
	\item[] Returns field name as it should be written in the constant \code{Braket} part.
\end{enumerate}
The convention to write a field in the constant part of a Braket is the following: for each field we assign a name, then we write the number of upper indices followed by the number of lower indices, then between parentheses we add the indices, the first index is always reserved for flavor if applicable, then we write the upper indices by the order they appear (left to right) followed by the lower indices (left to right). In the case we have some ambiguity concerning the symmetric or antisymmetric nature of the indices, we add the s letter for those that are symmetric just after the field name, e.g. the field ${M _{ij}}_b$ with symmetric $i$, $j$ indices and flavor index $b$, must be written in the Braket constant part as \codeb{Ms02(b,i,j)} and declared to FORM as \code{Field(M, 0, 2, SYM_WITH_FLAVOR)}.
	\item \code{void CallForm(Braket &exp, bool print=true, bool all=true, string newidlabel="j")} \\[2mm] Creates the input file for FORM, run the FORM program and returns the result to an output file and/or to the screen.
	\item \code{void setFormRenumber()}\\[2mm] Sets "renumber 1;" in FORM input file.
  This option is used to renumber indices in order to allow further simplifications.
  However, in large expressions this must be avoided since it increases the computational time in FORM.
  The best way to use it is simplify the expression with FORM with this option unset, and then send a second time to FORM
  with this option active. By default this option is unset.\\[2mm]
\code{void unsetFormRenumber()}\\[2mm] 
Unsets "renumber 1;" in FORM input file.
\item \code{void setFormIndexSum()}\\
\code{void unsetFormIndexSum()} \\[2mm] 
Sets/Unsets the index sum in input FORM file. The set option is activated by default.
\end{itemize}

%%%%%%%%%%%%%%%%%%%%%%%%%%%%%%%%%%%%%%
\section{Work with \texorpdfstring{\protect\SOSpin}{SOSpin}}
\label{sec:workprog}

In this section, we describe the installation of the \protect\SOSpin library and other tools that may be provided for the library to make further simplifications. We give in detail instructions how to use the \protect\SOSpin library in an standalone \Cpp program with a very simple example involving the group $\mathsf{SO}(4)$ just for illustration.

%%%%%%%%%%%%%%%%%%%%%%%%%%%%%%%%%%%%%%
\subsection{Download and installation}
\label{sec:inst} 

The \protect\SOSpin library project is hosted by Hepforge at \url{http://sospin.hepforge.org} under a GNU Lesser general public license. 

The simplest way to compile the \protect\SOSpin library is:
\begin{enumerate}
	\item \code{./configure --prefix=library_installation_path --with-form=FORM_path}\\[2mm]
	The user can omit the FORM path declaration and set it after using \code{export PATH_TO_FORM=FORM_path} or put the binary FORM file in the folder where the user has its project.
	\item \code{make}
	\item \code{make install}
	\item \code{make doxygen-doc} (optional - it generates \protect\SOSpin library documentation)
\end{enumerate}
Inside the library folder there are several example files, in addition to the ones shown in this paper, to help with the use of the library. The FORM~\cite{Kuipers:2012rf} binary files can be downloaded at \url{http://www.nikhef.nl/~form/}, after accepting the license agreement. They are available for LINUX (32-bits or 64-bits), Cygwin (32-bits) and Apple/Intel platforms. All the information concerning the installation is written in the README file. These procedures were successfully tested in LINUX and Mac OS X 10.10 (Yosemite).

%%%%%%%%%%%%%%%%%%%%%%%%%%%%%%%%%%%%%%
\subsection{Writing the first program}

In this section we discuss how to write a first program example using the \protect\SOSpin library properly for  $\mathsf{SO}(4)$. Although the group $\mathsf{SO}(4)\simeq \mathsf{SU}(2)\times \mathsf{SU}(2)$ is not particularly interesting for GUTs, it is invoked to illustrate the use of the library in a simpler way. Since $\mathsf{SO}(4)$ belongs to the family group $\mathsf{SO}(2N)$ for $N$ even, the two spinors in which the 4-dimensional space is broken are not related by conjugation. The general ket is given by,
\begin{equation}
\label{eq:genSO4}
	\Ket{\psi} = \Ket{0} M \,+\, b^{\dagger}_k \Ket{0} N^k \,+\, 
	\frac12\varepsilon^{ij}b^{\dagger}_ib^{\dagger}_j\Ket{0} \overline{M}\,,
\end{equation}
where $M,\overline{M}\sim \mathsf{1}$ and $N^i\sim \mathsf{2}$ in $\mathsf{SU}(2)$.
So, the general ket in \cref{eq:genSO4} can be decomposed as,
\begin{equation}
	\Ket{\psi} = \Ket{\psi_1} + \Ket{\psi_2}\,,
\end{equation}
where $\Gamma_0\Ket{\psi_1}=\Ket{\psi_1}$ and $\Gamma_0\Ket{\psi_2}=-\Ket{\psi_2}$, which are written as
\begin{align}
	\label{eq:k1}
	\Ket{\psi_1} &= \Ket{0}M + \frac12\varepsilon^{ij}b^{\dagger}_ib^{\dagger}_j\Ket{0} \overline{M}\,,\\
		\label{eq:k2}
	\Ket{\psi_2} &= b^{\dagger}_k \Ket{0}N^k \,.
	\end{align}
Using the rules compiled in \cref{subsec:comp} for the transposition operations, we obtain the following expressions
\begin{align}
\label{eq:b1}
\Bra{\psi^{\ast}_1}&=M\Bra{0} + \frac12\varepsilon^{lm}\overline{M}\Bra{0} b_m b_l\,,\\
\label{eq:b2}
\Bra{\psi^{\ast}_2}&=N^n\Bra{0}b_n\,.
\end{align}

As our first example program, we will address the calculation of $\Braket{\psi^{\ast}_1|B|\psi_1}$. In order to better understand how the library works, let us start by showing first how to code $\Braket{\psi^{\ast}_1|\psi_1}$ ignoring the fields and taking into account only the creation and annihilation operators, i.e.,
\begin{equation}
\label{eq:eqso4firstexample}
	\Big(\Bra{0} + \Bra{0} b_m b_l \Big) \Big(\Ket{0} + b^{\dagger}_ib^{\dagger}_j\Ket{0}\Big)\,.
\end{equation}
To properly write the code for this expression, we first need to add the header file for the \protect\SOSpin library and the sospin \code{namespace}. After creating the main function we must define the group dimension in the beginning, \code{setDim(4)}, and clean up all the memory allocated using \code{CleanGlobalDecl()} before exiting from the program. The code to solve the problem in \cref{eq:eqso4firstexample} is the following
\begin{lstlisting}
#include <sospin/son.h>
using namespace sospin;

int main(int argc, char *argv[]){
 setDim(4);
 Braket left = identity;
 left += b(m)*b(l);
 Braket right = identity;
 right += bt(i)*bt(j);
 Braket res = left * right;
 res.evaluate();
 
 res.setON();
 std::cout<<"Result:\n"<< res <<std::endl;
 res.setOFF();
 
 CleanGlobalDecl();
 
 exit(0);
}
\end{lstlisting}

The \code{evaluate()} function can be used with or without the arguments: true or false. 
The \code{evaluate(false)} sets on the results written in terms of the Levi-Civita and it is only used in operations with \code{braket} type; \code{evaluate()}/\code{evaluate(true)} does the evaluation to delta functions and it can be used in \code{bra()}/\code{ket()}/\code{braket()} and \code{none} types. If we use more than one evaluation process, we need to maintain the type of evaluation chosen.

The functions \code{setON()} and \code{setOFF()} make possible the writing of \code{Local R?=} for each term in \code{Braket} expressions, in addition, each term of the expression is numbered. Note that in the code written above, due to the way how it was declared, the terms need to be joined by using the += operation.

In order to compile and run the program, the user must pass to the \Cpp compiler the path to the \protect\SOSpin library and include folder, as for example (assuming the GCC compiler),
\begin{lstlisting}
 g++ -O3 -I/Sospin_PATH/include -L/Sospin_PATH/lib example.cpp -o example -lsospin 
\end{lstlisting}
Running the program above, we will get the following result\footnote{Note that, in order to save space, we altered slightly the aspect of the program's output.},
\begin{lstlisting}
Local R1 = +1;
Local R2 = + bt(i) * bt(j);
Local R3 = + b(m) * b(l);
Local R4 = + d_(m,j) * d_(l,i)
	     - d_(m,i) * d_(l,j)
	     - d_(l,i) * bt(j) * b(m)
	     + b(m) * bt(i) * bt(j) * b(l)
	     + d_(l,j) * bt(i) * b(m);
\end{lstlisting}

This result is not in agreement with the rules given in \cref{sec:prog}. Following those rules and looking at the term above we see that all the terms containing operators must vanish. The reason why these terms appear in the result above is because we never declared the type of \code{left}, \code{right} and \code{res}, hence by default all these expressions are of type \code{free} (operation \code{none}). In order to properly solve \cref{eq:eqso4firstexample} one needs to setup explicitly the type of each expression; we can declare them as
\begin{lstlisting}
left.Type() = bra;
right.Type() = ket;
\end{lstlisting}

There is no need to declare the variable \code{res} because the operation \code{left*right} will automatically setup its type based on the product, i.e., the resulting type of \code{res} is \code{braket}.

A simpler and more complete way to declare the expressions in \cref{eq:eqso4firstexample} is the following,
\begin{lstlisting}
 Braket left = Braket(identity, bra);
 left += Braket(b(m)*b(l), bra);
 Braket right = Braket(identity, ket);
 right += Braket(bt(i)*bt(j), ket);
\end{lstlisting}
where we use the operation \code{+=} for each contribution in different lines or 
\begin{lstlisting}
 Braket left = Braket(identity, bra) + Braket(b(m)*b(l), bra);
 Braket right = Braket(identity, ket) + Braket(bt(i)*bt(j), ket);
\end{lstlisting}
where we use the \code{+} operation for terms placed in the same line.

Running the program with the types properly setup, we get the expected result:
\begin{lstlisting}
Local R1 = + 1;
Local R2 = + d_(m,j) * d_(l,i)
	     - d_(m,i) * d_(l,j);
\end{lstlisting}
If we wish to evaluate the expression in~\cref{eq:eqso4firstexample} in such a way that the final result appears written in terms of the Levi-Civita tensors, we need simply set the argument of the \code{evaluate} function to false as \code{res.evaluate(false)}. The result will be given by the output
\begin{lstlisting}
Local R1 = +1;
Local R2 = +e_(m,l)*e_(j,i);
\end{lstlisting}
As one can see, the final result written in terms of the Levi-Civita tensor is equal to the one written in terms of deltas but in a much more compact form. So, hereinafter we will just present the results written in terms of Levi-Civita tensors even though both methods are available.

Let us now discuss how to include in \cref{eq:eqso4firstexample} the operator $B$ of \cref{eq:B}, i.e.,
\begin{equation}
	\Big(\Bra{0} + \Bra{0} b_m b_l \Big)\, B\, \Big(\Ket{0} + b^{\dagger}_ib^{\dagger}_j\Ket{0}\Big)\,.
\end{equation}
Writing this code is rather simple because the \SOSpin library already have a function to compute the operator $B$ for any group $\mathsf{SO}(2N)$, \code{Bop()}, therefore we only need to add this function to the code as:
\begin{lstlisting}
 Braket res = left * Bop() * right;
\end{lstlisting}
There is no need to set the group dimension in \code{Bop()} function because it is already done through the \code{setDim()} in the beginning of the code. If one wants to define the operator $B$ by oneself, without using the predefined function, we can do it just by using the \code{free()} function and the rules given above. The result will be
\begin{lstlisting}
Local R1 = 1/2*e_(i1,i2)*( -e_(i1,i2)*e_(j,i) );
Local R2 = 1/2*e_(i1,i2)*( +e_(i1,t1)*e_(i2,t1) );
Local R3 = 1/2*e_(i1,i2)*(
	     +d_(i1,i2)*e_(m,l)*e_(j,i)
	     -d_(i1,i)*e_(m,l)*e_(j,i2)
	     +d_(i1,j)*e_(m,l)*e_(i,i2)
);
Local R4 = 1/2*e_(i1,i2)*(
	     +d_(i2,i)*e_(m,l)*e_(j,i1)
	     -d_(i2,j)*e_(m,l)*e_(i,i1)
);
Local R5 = 1/2*e_(i1,i2)*( -e_(m,l)*e_(i2,i1) );
\end{lstlisting}

These results are quite large and clumsy so in order to simplify them we decided to include FORM as a final step. The inclusion of FORM is purely aesthetics and does not affect the computation procedures. In the case we consider it the last result become
\begin{lstlisting}
Local R1 = +e_(i,j);
Local R2 = -e_(l,m);
\end{lstlisting}

After this introduction, we are ready to compute $\Braket{\psi_{1a}^{\ast}|B|\psi_{1b}}$, i.e.,
\begin{equation}
	\Big(M_a\Bra{0} + \frac12\varepsilon^{lm}\overline{M}_a\Bra{0} b_m b_l\Big)\, B\, \Big(\Ket{0}M_b + \frac12\varepsilon^{ij}b^{\dagger}_ib^{\dagger}_j\Ket{0} \overline{M}_b\Big)\,,
\end{equation}
where $a$ and $b$ are flavor indices.

In order to compute this example, we need to add the fields $M$ and $\overline{M}$. This is done by using the functions \code{bra()}, \code{ket()}, \code{free()} and \code{braket()} defined in \cref{sec:generalfunctions}. To account for the changes due to the inclusion of the fields we need to substitute the codes given above by
\begin{lstlisting}
 Braket left = bra(0,M(a),identity);
 left +=bra(0,Mb(a)*e_(l,m)/2,b(m)*b(l));
 Braket right = ket(0,M(b),identity);
 right +=ket(0,Mb(b)*e_(i,j)/2,bt(i)*bt(j));
\end{lstlisting}
where the field $\overline{M}$ is coded as \code{Mb}. The output result is 
\begin{lstlisting}
Local R1 = M(a)*1/2*e_(i1,i2)*Mb(b)*e_(i,j)/2*(-e_(i1,i2)*e_(j,i));
Local R2 = M(a)*1/2*e_(i1,i2)*M(b)*( +e_(i1,t1)*e_(i2,t1) );
Local R3 = Mb(a)*e_(l,m)/2*1/2*e_(i1,i2)*Mb(b)*e_(i,j)/2*(
	     +d_(i1,i2)*e_(m,l)*e_(j,i)
	     -d_(i1,i)*e_(m,l)*e_(j,i2)
	     +d_(i1,j)*e_(m,l)*e_(i,i2)
 );
Local R4 = Mb(a)*e_(l,m)/2*1/2*e_(i1,i2)*Mb(b)*e_(i,j)/2*(
	     +d_(i2,i)*e_(m,l)*e_(j,i1)
	     -d_(i2,j)*e_(m,l)*e_(i,i1)
 );
Local R5 = Mb(a)*e_(l,m)/2*1/2*e_(i1,i2)*M(b)*( -e_(m,l)*e_(i2,i1) );
\end{lstlisting}

To make the final simplification we use the FORM program; before we call it to simplify our expression, we first need to declare explicitly all fields as well as all indices appearing only in the constant part, i.e., the indices $a$ and $b$ in this example. Therefore, we need to add the following code,
\begin{lstlisting}
 Field(M, 0, 0, ASYM_WITH_FLAVOR);
 Field(Mb, 0, 0, ASYM_WITH_FLAVOR);
 newId("a"); newId("b"); 
\end{lstlisting}

Once the fields $M$ and $\overline{M}$ carry flavor, we need to set them as \code{ASYM_WITH_FLAVOR}, for more details see \cref{sec:generalfunctions}. 
To call the FORM program to simplify our expression we only need to write
\begin{lstlisting}
 CallForm(res,false, true, "j");
\end{lstlisting}
Note that the function \code{callForm()} sets \code{setOFF()} for the expression \code{Braket}. If the user have declared \code{setON()} previously, the user must set \code{setON()} again for that expression after \code{callForm()} . For a more detailed description about this function please see \cref{sec:generalfunctions}.

Running the program above we obtain the following result,
\begin{lstlisting}
Local R1 = + M(j1) * Mb(j2);
Local R2 = - Mb(j1) * M(j2);
\end{lstlisting}
Note that j1 and j2 are not arguments but indices; they correspond to the flavor indices a and b. This is so because by default we consider that the indices are summed and hence they are renamed. If we want to avoid summed indices we must add the function \code{unsetFormIndexSum()} and then the result would be
\begin{lstlisting}[deleteemph={b}]
Local R1 = + M(a) * Mb(b);
Local R2 = - Mb(a) * M(b);
\end{lstlisting}

For the sake of completeness, we give below the complete code to compute $\Braket{\psi_{1a}^{\ast}|B|\psi_{1b}}$ in $\mathsf{SO}(4)$.
\begin{lstlisting}
#include <sospin/son.h>
using namespace sospin;

int main(int argc, char *argv[]){

 setDim(4);

 Braket left = bra(0,M(a),identity);
 left +=bra(0,Mb(a)*e_(l,m)/2,b(m)*b(l));
 Braket right = ket(0,M(b),identity);
 right +=ket(0,Mb(b)*e_(i,j)/2,bt(i)*bt(j)); 

 Braket res = left * Bop() * right;
 res.evaluate(false);

 Field(M, 0, 0, ASYM_WITH_FLAVOR);
 Field(Mb, 0, 0, ASYM_WITH_FLAVOR);
 newId("a"); newId("b"); 

 unsetFormIndexSum(); 
 
 CallForm(res,false, true, "j");
 
 res.setON();
 std::cout<<"Result:\n"<< res <<std::endl;
 res.setOFF(); 
 
 CleanGlobalDecl();
 exit(0);
}
\end{lstlisting}
where the result is obviously $M_a\,\overline{M}_b - \overline{M}_a\, M_b\,$.
In order to compute $\Braket{\psi_{1a}^{\ast}|B|\psi_{2b}}$, $\Braket{\psi_2^{\ast}|B|\psi_1}$ and $\Braket{\psi_{2a}^{\ast}|B|\psi_{2b}}$ we just need to define $\Bra{\psi^\ast_{2a}}$ and $\Ket{\psi_{2b}}$, and substitute it in the code above. Using \cref{eq:b2} we define $\Bra{\psi^\ast_{2a}}$ as 
\begin{lstlisting}
Braket left = bra(1,N(a,n),b(n));
\end{lstlisting}
and using \cref{eq:k2} we define $\Ket{\psi_{2b}}$ as 
\begin{lstlisting}
Braket right = ket(1,N(b,k),bt(k));
\end{lstlisting}

The results are:
\begin{equation}
	\begin{aligned}
	\Braket{\psi_{1a}^{\ast}|B|\psi_{2b}} &=\Braket{\psi_2^{\ast}|B|\psi_1}=\,0\,,\\
	\Braket{\psi_{2a}^{\ast}|B|\psi_{2b}} &=\varepsilon^{ij} N_a^iN_b^{\prime j}\,.
	\end{aligned}
\end{equation}

\begin{table*}
\caption{\label{tab:so4_1a1b} The \protect\SOSpin code to compute $\Braket{\psi_{1a}^{\ast}|B\Gamma_{\mu}|\psi_{2b}} \Braket{\psi_{1c}^{\ast}|B\Gamma_{\mu}|\psi_{2d}}$ in $\mathsf{SO}(4)$ and the corresponding result. In this example we have used the operator $B$, \code{Bop()}, that uses generic indices, i and k.}
\begin{center}
\begin{tabular}{p{\textwidth}}
\hline
\begin{minipage}{\textwidth}
\protect{\begin{lstlisting}[basicstyle=\small,numbers=left,deleteemph={b,bt}]
#include <sospin/son.h>
using namespace sospin;

int main(int argc, char *argv[]){

 setDim(4);

 Braket L1, R1, L2, R2;
 Braket in1E, in1O, in2E, in2O, res;

 L1 = bra(0,M(a),identity);
 L1+= bra(0,Mb(a)*e_(i,j)/2,b(j)*b(i));
 R1 = ket(0,N10(b,k),bt(k));

 L2 = bra(0,M(c),identity);
 L2+= bra(0,Mb(c)*e_(l,m)/2,b(m)*b(l));
 R2 = ket(0,N10(d,o),bt(o));

 Field(M, 0, 0, ASYM_WITH_FLAVOR);
 Field(Mb, 0, 0, ASYM_WITH_FLAVOR);
 Field(N, 1, 0, ASYM_WITH_FLAVOR);

 newId("a"); newId("b");
 newId("c"); newId("d");

 in1E = L1 * Bop("i") * G(true, "j") * R1;
 in2E = L2 * Bop("k") * G(true, "j") * R2;
 in1O = L1 * Bop("i") * G(false,"j") * R1;
 in2O = L2 * Bop("k") * G(false,"j") * R2;

 in1E.evaluate();
 in2E.evaluate();
 in1O.evaluate();
 in2O.evaluate();

 res = in1E * in2E + in1O * in2O ;

 unsetFormIndexSum();
 CallForm(res,false, true, "i");

 res.setON();
 std::cout << "Output result:\n" << res << std::endl;

 CleanGlobalDecl();

 exit(0);
}
\end{lstlisting}}
\end{minipage}
\\\hline \\[-1mm]
Output result:\\
\hspace{1cm}\begin{minipage}{\textwidth}
\protect{\begin{lstlisting}[numbers=none,deleteemph={b,bt}]
Local R1 = +2*M(a)*N10(b,i)*Mb(c)*N10(d,j)*e_(i,j);
Local R2 = -2*Mb(a)*N10(b,i)*M(c)*N10(d,j)*e_(i,j); 
\end{lstlisting}}
\end{minipage}
\\
\hline
\end{tabular} 
\end{center}
\end{table*}
	
%%%%%%%%%%%%%%%%%%%%%%%%%%%%%%%%%%%%%%
\section{Examples}
\label{sec:examples}

In this section we present two more complex examples to better explain how to use \SOSpin library. We give one example in $\mathsf{SO}(4)$ with higher dimensional terms and one example in the context of $\mathsf{SO}(10)$ models. 

%%%%%%%%%%%%%%%%%%%%%%%%%%%%%%%%%%%%%%
\subsection{$\mathsf{SO}(4)$}

Let us compute the following higher dimensional interaction term in $\mathsf{SO}(4)$:
\begin{equation}
\label{eq:higher}
\frac{Y_{ab} Y_{cd}}{\Lambda^2} \sum^4_{\mu=1}\Braket{\psi_{1a}^{\ast}|B\Gamma_{\mu}|\psi_{2b}} \Braket{\psi_{1c}^{\ast}|B\Gamma_{\mu}|\psi_{2d}}\,,
\end{equation}
where $Y_{ab}$, $Y_{cd}$ are Yukawa matrices, $a$, $b$, $c$, $d$ are flavor indices and $\Lambda$ is some high energy scale. The above equation can be expanded by rewriting the $\Gamma_\mu$ in terms of the operators $b_i$ and $b^{\dagger}_i$ as given in \cref{eq:gammamatrices}
\begin{equation}
\begin{aligned}
&\frac{Y_{ab} Y_{cd}}{\Lambda^2} \sum^2_{i=1}\biggl(
\Braket{\psi_{1a}^{\ast}|B(b_i+b^{\dagger}_i)|\psi_{2b}} \Braket{\psi_{1c}^{\ast}|B(b_i+b^{\dagger}_i)|\psi_{2d}}\biggr. \\
&\biggl.\,-\,\Braket{\psi_{1a}^{\ast}|B(b_i-b^{\dagger}_i)|\psi_{2b}} \Braket{\psi_{1c}^{\ast}|B(b_i-b^{\dagger}_i)|\psi_{2d}}
\biggr)\,.
\end{aligned}
\end{equation}

The code to compute \cref{eq:higher} is given in \cref{tab:so4_1a1b} and the corresponding $\Gamma_{\mu}$ is given by taking into account the parity as
\begin{lstlisting}
Braket G(bool even, string startid){
 if(even){
  Braket G_even = bb(startid +"1");
  G_even += bbt(startid +"1");
  return G_even;
 }
 Braket G_odd = bb(startid +"2");
 G_odd -= bbt(startid +"2");
 string constpart = "-i_";
 G_odd = G_odd * constpart;
 return G_odd;
}
\end{lstlisting}
Note that it is important to carefully define different indices among bracket expressions in order to avoid repeated indices, which could lead to a meaningless result. The result of the above code is then given as 
\begin{equation}
2\frac{Y_{ab} Y_{cd}}{\Lambda^2}\left(M_a N^i_b \overline{M}_c N^j_d - \overline{M}_a N^i_b M_c N^j_d \right) \varepsilon_{ij}\,.
\end{equation}

%%%%%%%%%%%%%%%%%%%%%%%%%%%%%%%%%%%%%%
\subsection{$\mathsf{SO}(10)$} 
\label{sec:so10}

In this subsection, we give the example $\mathsf{16}_a\,\mathsf{16}_b\,\mathsf{120}_H$ of $\mathsf{SO}(10)$ computed using the \SOSpin library. In a first step we present the program by defining all quantities while in a second step we rewrite it using only the specific $\mathsf{SO}(10)$ functions already included in the library. The reducible $\mathsf{32}$ representation of $\mathsf{SO}(10)$ is given by
\begin{equation}
\begin{aligned}
\Ket{\Psi}=&\Ket{0}\psi + b^{\dagger}_i\Ket{0}\psi^i+\frac{1}{2}b^{\dagger}_ib^{\dagger}_j\Ket{0}\psi^{ij}
+ \frac{1}{12}\varepsilon^{ijklm}b^{\dagger}_k b^{\dagger}_l b^{\dagger}_m \Ket{0}\overline{\psi}_{ij}\\
&+ \frac{1}{24}\varepsilon^{ijklm}b^{\dagger}_j b^{\dagger}_k b^{\dagger}_l b^{\dagger}_m \Ket{0}\overline{\psi}_{i}
+b^{\dagger}_1 b^{\dagger}_2 b^{\dagger}_3 b^{\dagger}_4 b^{\dagger}_5 \Ket{0}\overline{\psi}\,.
\end{aligned}
\end{equation}
As already mentioned, in $\mathsf{SO}(10)$ the fermionic particles are usually assigned to the $\mathsf{16}$ dimensional representation which corresponds to the semi-spinor $\Psi_{+}$ while $\overline{\mathsf{16}}\equiv \Psi_{-}$. The spinor representations are schematically given in \cref{tab:spinor} where $\ket{\Psi_+}$ and $\ket{\Psi_-}$ are given by
\begin{equation}
\label{eq:psi+}
\Ket{\Psi_+}=\Ket{0} M+ \frac{1}{2}b^{\dagger}_jb^{\dagger}_k\Ket{0}M^{jk}+\frac{1}{24}\varepsilon^{jklmn}b^{\dagger}_k b^{\dagger}_l b^{\dagger}_m b^{\dagger}_n \Ket{0}\overline{M}_{j}\,,
\end{equation}
and 
\begin{equation}
\label{eq:psi-}
\Ket{\Psi_{-}}=
b^{\dagger}_i\Ket{0}\,M^i
+ \frac{1}{12}\varepsilon^{ijklm}\,b^{\dagger}_k b^{\dagger}_l b^{\dagger}_m \Ket{0}\,\overline{M}_{ij}
+ b^{\dagger}_1 b^{\dagger}_2 b^{\dagger}_3 b^{\dagger}_4 b^{\dagger}_5 \Ket{0}\,\overline{M}\,,
\end{equation}
while the transpose of $\Ket{\Psi}$, represented as $\Bra{\Psi^{\ast}}$ are given by,
\begin{equation}
\label{eq:plusconj}
\Bra{\Psi^{\ast}_{+}}=M\Bra{0} 
+ \frac{1}{2}M^{ij}\Bra{0}b_jb_i
+ \frac{1}{24}\varepsilon^{ijklm}\,\overline{M}_{m}\Bra{0}b_l b_k b_j b_i\,,
\end{equation}
and
\begin{equation}
\label{eq:bram}
\Bra{\Psi^{\ast}_{-}}=M^i \Bra{0} b_i
+ \frac{1}{12}\varepsilon^{ijklm}\,\overline{M}_{ij}\Bra{0}b_m b_l b_k 
+ \overline{M}\Bra{0}b_5 b_4 b_3 b_2 b_1 \,,
\end{equation}
where the flavor index was omitted.

\begin{table}
	\begin{center}
		\begin{tabular}{ccc}
			\hline
			$\mathsf{SO}(10)$ & \multicolumn{2}{c}{$\mathsf{SU}(5)$}\\
			\hline
			$\Ket{0}$ & 1 & $M$ \\[3mm]
			$b^{\dagger}_j\Ket{0}$ & 5 & $M^{i}$\\[3mm]
			$b^{\dagger}_jb^{\dagger}_k\Ket{0}$ & 10 & $M^{ij}$ \\[3mm]
			$b^{\dagger}_jb^{\dagger}_k b^{\dagger}_l \Ket{0}$ & $\overline{10}$ & $\overline{M}_{ij}$ \\[3mm]
			$b^{\dagger}_jb^{\dagger}_k b^{\dagger}_l b^{\dagger}_m \Ket{0}$ & $\overline{5}$ & $\overline{M}_i$\\[3mm]
			$b^{\dagger}_1b^{\dagger}_2 b^{\dagger}_3 b^{\dagger}_4 b^{\dagger}_5 \Ket{0}$ & 1 & $\overline{M}$\\[3mm]
			\hline
			& $\rm{dim}=2^5=32$
		\end{tabular}
		\caption{\label{tab:spinor} The $\mathsf{SO}(10)$ states in terms of $\mathsf{SU}(5)$ fields.}
	\end{center}
\end{table}

\begin{table*}
\caption{\label{tab:T120} The \protect\SOSpin code to compute $\mathsf{16}_a\,\mathsf{16}_b\,\mathsf{120}_H$, i.e., $\frac{1}{3!}\Bra{\Psi^{\ast}_{+\,a}}B\ \Gamma_{\mu}\Gamma_{\nu}\Gamma_{\rho}\Ket{\Psi_{+\,b}}\Phi_{\mu\nu\rho}$ in $\mathsf{SO}(10)$ and the correspondig result.}
\begin{center}
\begin{tabular}{p{\textwidth}}
\hline
\begin{minipage}{\textwidth}
\lstset{escapeinside={(*@}{@*)}}
\protect{\begin{lstlisting}[basicstyle=\small,numbers=left,deleteemph={b,bt}]
#include <sospin/son.h> (*@\label{L1}@*) 
using namespace sospin;

int main(int argc, char *argv[]){
 setDim(10); (*@\label{L2}@*) 

 Braket psipbra = bra(0, M(a), identity);(*@\label{L3}@*) 
 psipbra += bra(2, 1/2*M20(a,o,p),-b(o)*b(p)); 
 psipbra += bra(4,1/24*e_(o,p,q,r,s)*Mb01(a,o),b(p)*b(q)*b(r)*b(s));  
  
 Braket psipket = ket(0,M(b), identity); (*@\label{L4}@*) 
 psipket += ket(2,1/2 * M20(b,j,k), bt(j) * bt(k));
 psipket += ket(4, 1/24*e_(j,k,l,m,n)*Mb01(b,j), bt(k)*bt(l)*bt(m)*bt(n)); 
 
 newId("a");   newId("b"); (*@\label{L5}@*)
 Field(M, 0, 0, ASYM_WITH_FLAVOR); (*@\label{L6}@*)
 Field(M, 2, 0, ASYM_WITH_FLAVOR);
 Field(Mb, 0, 1, ASYM_WITH_FLAVOR);

 Braket gamma = free(-3,1/6*(e_(r1,r2,r3,r4,r5)*H20(r4,r5)/sqrt(3)),b(r1)*b(r2)*b(r3));(*@\label{L7}@*)
 gamma += free(3,1/6*(e_(r1,r2,r3,r4,r5)*H02(r4,r5)/sqrt(3)),bt(r1)*bt(r2)*bt(r3));
 gamma += free(-1,(2*H12(r1,r2,r3)+d_(r1,r2)*H01(r3)-d_(r1,r3)*H01(r2))/(2*sqrt(3)), bt(r1)*b(r2)*b(r3));
 gamma += free(1,(2*H21(r1,r2,r3)+d_(r1,r3)*H10(r2)-d_(r2,r3)*H10(r1))/(2*sqrt(3)), bt(r1)*bt(r2)*b(r3));
 gamma += free(-1,2*H01(r1)/sqrt(3),-b(r1));
 gamma += free(1,2*H10(r1)/sqrt(3), bt(r1));
 newId("r4");   newId("r5");  (*@\label{L8}@*) 

 Field(H, 1, 0, ASYM);   Field(H, 0, 1, ASYM); (*@\label{L9}@*) 
 Field(H, 0, 2, ASYM);   Field(H, 2, 0, ASYM);
 Field(H, 2, 1, ASYM);   Field(H, 1, 2, ASYM);

 Braket exp = psipbra * Bop() * gamma * psipket; (*@\label{L10}@*) 
 exp.evaluate(); 
 CallForm(exp,true, true); (*@\label{L11}@*) 

 CleanGlobalDecl(); (*@\label{L12}@*) 
 exit(0);
}
\end{lstlisting}}
\end{minipage}
\\\hline \\[-1mm]
Output result:\\
\hspace{1cm}\begin{minipage}{\textwidth}
\protect{\begin{lstlisting}[numbers=none,deleteemph={b,bt}]
R = + 2*M(j1)*H10(j2)*Mb01(j3,j2)*sqrt(1/3)*i_
  - M(j1)*H02(j2,j3)*M20(j4,j2,j3)*sqrt(1/3)*i_
  + M20(j1,j2,j3)*H01(j3)*Mb01(j4,j2)*sqrt(1/3)*i_
  + M20(j1,j2,j3)*H02(j2,j3)*M(j4)*sqrt(1/3)*i_
  + 1/4*M20(j1,j2,j3)*H21(j4,j5,j6)*M20(j7,j6,j8)*sqrt(1/3)*e_(j2,j3,j4,j5,j8)*i_
  - 1/4*M20(j1,j2,j3)*H21(j4,j5,j6)*M20(j7,j8,j6)*sqrt(1/3)*e_(j2,j3,j4,j5,j8)*i_
  - M20(j1,j2,j3)*H12(j4,j2,j3)*Mb01(j5,j4)*sqrt(1/3)*i_
  - 2*Mb01(j1,j2)*H10(j2)*M(j3)*sqrt(1/3)*i_
  - 1/2*Mb01(j1,j2)*H01(j3)*M20(j4,j2,j3)*sqrt(1/3)*i_
  + 1/2*Mb01(j1,j2)*H01(j3)*M20(j4,j3,j2)*sqrt(1/3)*i_
  + 2*Mb01(j1,j2)*H20(j2,j3)*Mb01(j4,j3)*sqrt(1/3)*i_
  + Mb01(j1,j2)*H12(j2,j3,j4)*M20(j5,j3,j4)*sqrt(1/3)*i_;
\end{lstlisting}}
\end{minipage}
\\
\hline
\end{tabular} 
\end{center}
\end{table*}

The Yukawa term $Y_{ab}\,\mathsf{16}_a\,\mathsf{16}_b\,\mathsf{120}_H$ is written as 
\begin{equation}
\frac{1}{3!}Y_{ab}\Bra{\Psi^{\ast}_{+\,a}}B\ \Gamma_{\mu}\Gamma_{\nu}\Gamma_{\rho}\Ket{\Psi_{+\,b}}\Phi_{\mu\nu\rho}\,.
\end{equation}

In order to compute this expression using the \SOSpin library, we write our main program as done before for $\mathsf{SO}(4)$ examples. The code to compute this example is given in~\cref{tab:T120}. We start by including the \code{son.h} file, \mbox{\code{sospin/son.h}} in line~\ref{L1}, then we set the group dimension with the function \mbox{\code{setDim(10)}} (line~\ref{L2}). The expression for $\Ket{\Psi_+}$ is given in~\cref{eq:psi+} and is declared in line~\ref{L4}; $\Bra{\Psi^{\ast}_{+}}$ is given in~\cref{eq:plusconj} and declared in line~\ref{L3} while the flavor indices $a$ and $b$ are declared in line~\ref{L5} in~\cref{tab:T120}.

Making use of the basic theorem~\cite{Nath:2001yj} pointed out in \cref{eq:basictheo}, one can write the action of $\Gamma_\mu \Gamma_\nu \Gamma_\rho$ over the 120-dimensional Higgs field $\Phi_{\mu\nu\rho}$ which is given in \cref{eq:gen120new} of \cref{sec:appSO10} and defined in line~\ref{L7} in~\cref{tab:T120}. The charge conjugation operator, $B$, is in this code taken as the internal function \code{Bop()} (defined in \cref{sec:generalfunctions}). In line~\ref{L8} we declare the indices that appear in \code{gamma} but are not defined automatically (i.e., the ones that appear only in fields, r4 and r5 in this case). The definition of fields in FORM language is done in lines~\ref{L6} and~\ref{L9}. In line~\ref{L10} the expression is evaluated, in line~\ref{L11} we call the FORM to perform some final simplifications and in line~\ref{L12} we clean temporary allocated memory with \code{CleanGlobalDecl()}.

Note that $Y_{ab}$ carries flavor dependency, but it is not written explicitly in the code, and hence the final result presented in \cref{tab:T120} can be simplified by performing the symmetrisation of the fields using
\begin{equation}
A_i\,B_j=\frac{1}{2}(A_i\,B_j+A_j\,B_i) +\frac{1}{2}(A_i\,B_j-A_j\,B_i).
\end{equation}
If we work out the final result of \cref{tab:T120} by using the expression above, we obtain
\begin{equation}
\label{eq:base4}
\begin{aligned}
i\,\frac{2}{\sqrt{3}}\,Y^-_{ab}\,&\left( 2\,M_a\overline{M}_b\,H^i\,+\,
M^{ij}_a\,M_b\,H_{ij}\right.\\
&\,+\,\overline{M}_{i\,a}\,\overline{M}_{j\,b}\,H^{ij}\,-\,
M^{ij}_a\,M_b\,H_j\\
&\left.\,+\,
\overline{M}_{i\,a}\,M^{jk}_b\,H^i_{jk}\,-\,
\frac{1}{4}\,\varepsilon_{ijklm}\,M^{ij}_a\,M^{mn}_b\,H^{kl}_n
\right)\,,
\end{aligned}
\end{equation}
where $Y^-_{ab}\equiv\dfrac{Y_{ab}-Y_{ba}}{2}$.

Now we write down the specific function of $\mathsf{SO}(10)$ that are implemented in \SOSpin library; these functions are defined in \code{tools/so10.h} file.

\begin{itemize} 
	\item	\code{Braket GammaH(int n)}\\[2mm]
Gives the action of the \code{n} $\Gamma$-matrices acting over the Higgs field, as summarised in \cref{sec:appSO10}; \code{n} is the number of $\Gamma$-matrices (runs from 0 to 5).
	
	\item	\code{Braket psi\_16p(OPMode mode, string id)}\\
	\code{Braket psi\_16p(OPMode mode)}\\[2mm]
	These functions give $\Bra{\psi^\ast_+}$, defined in \cref{eq:plusconj}, if the mode is bra and $\Ket{\psi_+}$, defined in \cref{eq:psi+}, if we select the ket mode. 
	
	\item \code{Braket psi\_16m(OPMode mode, string id)}\\
	\code{Braket psi\_16m(OPMode mode)}\\[2mm]
	These functions give $\Bra{\psi^\ast_-}$, defined in \cref{eq:bram}, if the mode is bra and $\Ket{\psi_-}$, defined in \cref{eq:psi-} if the selected mode is ket.
	
	\item	\code{Braket psi\_144p(OPMode mode)}\\
	\code{Braket psi\_144m(OPMode mode)}\\[2mm]
	Functions for $144_+$ and $144_-$, defined for modes bra and ket. The complete expressions are given in Ref.~\cite{Nath:2005bx}.
\end{itemize}

In what follows, we rewrite the code presented in \cref{tab:T120} just making use of the included functions described above. 

\begin{lstlisting}[breaklines=false]
#include <sospin/son.h>
#include <sospin/tools/so10.h>

using namespace sospin;

int main(int argc, char *argv[]){
	
	setDim(10);
	
	Braket res = psi_16p(bra) * Bop() * 
	       GammaH(3) * psi_16p(ket);
	
	res.evaluate(); 	
	CallForm(res, true, true); 
	
	CleanGlobalDecl();
	exit(0);
}
\end{lstlisting}

%%%%%%%%%%%%%%%%%%%%%%%%%%%%%%%%%%%%%%
\section{Conclusions}
\label{sec:conclusions}

In this paper we have presented the \SOSpin library, provided under the terms of the GNU Lesser General Public License as published by the Free Software Foundation. This is a \Cpp tool whose main goal is to decompose Yukawa interactions, invariant under $\mathsf{SO}(2N)$, in terms of $\mathsf{SU}(N)$ fields. The library project is hosted by the Hepforge website~\url{http://sospin.hepforge.org}.

This library relies on the oscillator expansion formalism that consists in expressing the $\mathsf{SO(2N)}$ spinor representations in terms of creation operators, $b^\dagger_i$, of a Grassmann algebra, acting on a vacuum state-vector. The \SOSpin code simulates the non-commutativity of the operators and their products via the implementation of doubly-linked-list data structures. Such type of structures are the ideal method to deal with the usage of long chains of products of operators $b^\dagger_i$ and $b_i$. In this type of implementation, the sequences are linked through \textit{nodes} that contains information about the previous and the next nodes of the sequence; this connections are called \textit{links}. Moreover, the data storage in the memory does not need to be adjacent, this is one of the reason why the doubly-linked-lists led to high performances in our tests.

In order to understand the manipulations that \SOSpin need to perform, we reviewed in detail the oscillator expansion formalism. Then, we applied the method for decomposing the $\mathsf{SO}(2N)$ Yukawa terms with respect to $\mathsf{SU}(N)$ interactions. The general structure of \SOSpin library was presented by listing the generic and devoted $\mathsf{SO}(10)$ functions. After explaining the installation and the writing of the first program, we showed in detail the usage of \SOSpin through complete examples in both $\mathsf{SO}(4)$ and $\mathsf{SO}(10)$ frameworks. Additionally, we provided an higher dimensional field-operator example computed in the context of $\mathsf{SO}(4)$ to illustrate how such a term can be processed with this library. Finally, we described the functions available in \SOSpin that were made to simplify the writing of spinors and their interactions specifically for $\mathsf{SO}(10)$ models. The code includes also functions to deal directly with the $\mathsf{144}$ and $\overline{\mathsf{144}}$ representations of $\mathsf{SO}(10)$ and one can compute other quantities beyond the results already computed in Ref.~\cite{Nath:2005bx}.

We are planning to enhance the use of the memory in our library by implementing new forms of simplifications. We intend to make the simplifications of the final expressions independent of external programs in order to reach more performance. Our future plans also include the extension of the library with specific functions, which are now only implemented for $\mathsf{SO}(10)$, automatised for a generic $\mathsf{SO}(2N)$. Furthermore, although the \SOSpin library was projected to cover the groups $\mathsf{SO}(2N)$, it is easily adapted to groups $\mathsf{SO}(2N+1)$ or other algebraic systems that are described by creation and annihilation operators. 

%%%%%%%%%%%%%%%%%%%%%%%%%%%%%%%%%%%%%%
\section*{Acknowledgments}
D.E.C. would like to thank Palash B. Pal for enlightening discussions. C.S. would like to thank Jean-René Cudell for useful comments.
D.E.C. and C.S. thank the Theory Unit of CERN Physics Department for the hospitality. D.E.C. also thanks Theory Unit of CERN for financial support.
The work of D.E.C.~is supported by Associa\c c\~ao do Instituto Superior T\'ecnico para a Investiga\c c\~ao e Desenvolvimento (IST-ID) and Funda\c{c}\~{a}o para a Ci\^{e}ncia e a Tecnologia (FCT) under the projects PTDC/FIS-NUC/0548/2012 and UID/FIS/00777/2013.
The work of C.S. is supported by the Universit\'e de Li\`ege and the EU in the context of the MSCA-COFUND-BeIPD project.

%%%%%%%%%%%%%%%%%%%%%%%%%%%%%%%%%%%%%%
\appendix

%%%%%%%%%%%%%%%%%%%%%%%%%%%%%%%%%%%%%%
\section{Clifford algebrae and $\mathsf{SO}(2N)$}
\label{sec:clifford}
In what follows we cover the Clifford algebra discussion given in Ref.~\cite{West:1998ey} and we adapt it for the specific algebra obtained from $\mathsf{SO}(2N)$. In general, we define the Clifford algebra as 
\begin{equation}
\label{eq:ca}
 \left\{\Gamma_{\mu},\Gamma_{\nu}\right\}\,=\,2\,\eta_{\mu}\delta_{\mu\nu}\,,
\end{equation}
where $\eta_{\mu}$ is a real constant with $|\eta_{\mu}|=1$. 
From \cref{eq:ca} one gets $\Gamma^2_{\mu}\,=\,\eta_{\mu}\mathbf{1}\,$ and
\begin{equation}
\label{eq:gammadagger}
\Gamma^{\dagger}_{\mu}\,=\, \eta_{\mu}\,\Gamma_{\mu}\,.
\end{equation}
For the spinor representation of $\mathsf{SO}(2N)$, the corresponding Clifford algebra has $\eta_{\mu}=1$ for any index $\mu$, i.e.,
\begin{equation}
\label{eq:ca2}
 \left\{\Gamma_{\mu},\Gamma_{\nu}\right\}\,=\,2\,\delta_{\mu\nu}\,.
\end{equation}
Moreover, one may construct a new independent matrix $\Gamma_0$, defined as
\begin{equation}
\Gamma_0\,\equiv\,i^N\Gamma_1\Gamma_2\cdots\Gamma_{2N}\,,
\end{equation}
which anticommutes with all $\Gamma$-matrices. This definition implies that $\Gamma^{\dagger}_{0}=\Gamma_{0}$, since
\begin{equation}
\begin{aligned}
\Gamma_{0}^{\dagger}\,&=\, (-i)^N {\Gamma_{2N}}^{\dagger}\cdots{\Gamma_1}^{\dagger} 
\,=\,
i^N (-1)^N \Gamma_{2N}\cdots\Gamma_1\,\\
&=\,i^N(-1)^{N(2N-1)+N} \Gamma_1\cdots\Gamma_{2N}\,=\,\Gamma_{0}\,,
\end{aligned}
\end{equation}
where one has used the following relation for non-repeating $p$ $\Gamma$-matrices with $p\leq 2N$:
\begin{equation}
\label{eq:invgamma}
\Gamma_{m_p}\cdots\Gamma_{m_2}\Gamma_{m_1}\,=\,(-1)^{\frac{p(p-1)}{2}}\Gamma_{m_1}\Gamma_{m_2}\cdots\Gamma_{m_p}\,.
\end{equation}
The exponent is obtained through the general formula for the sum of $p$ elements of an arithmetic progression $u_1,\dots,u_p$, i.e., $S_p=p\,\frac{u_1+u_p}{2}$. The first matrix $\Gamma_{m_p}$ of the right-handed side of \cref{eq:invgamma} has to anticommute $p-1$ times, the process ends with $\Gamma_1$, which does not need to move. One may then define the following projectors in analogy to the chiral projectors for fermion fields:
\begin{equation}
\label{eq:chiralprojectors}
P_L\,\equiv\,\frac{1\,-\,\Gamma_{0}}{2}\,,\quad
P_R\,\equiv\,\frac{1\,+\,\Gamma_{0}}{2}\,.
\end{equation}
They are indeed projectors, since
\begin{equation}
\begin{aligned}
&P_L\,+\,P_R\,=\,\mathbf{1}\,,\quad P_L^2\,=\,P_L\,,\quad P_R^2\,=\,P_R\,,\\
&P_L\,P_R\,=\,P_R\,P_L\,=\,0\,.
\end{aligned}
\end{equation}

In order to deduce the main results of this appendix, construct the set $C_D$ of all products of $\Gamma$-matrices of the Clifford algebra,
\begin{equation}
C_D\,=\,\bigl\{\pm1,\,\pm\,\Gamma_{m_1},\,\pm\,\Gamma_{m_1}\Gamma_{m_2},\,\dots,\pm\,
\Gamma_0\bigr\}\,,
\end{equation}
which is a finite group. Thus, we can enumerate all irreducible representations of the group $C_D$ compatible with the algebra. In particular, the unidimensional representations of $C_D$ cannot satisfy the anticommutation of \cref{eq:ca}. The order of the group $C_D$ is given by\begin{equation}
\ord(C_D)=\,2\sum_{i=0}^{2N}\binom{2N}{i}\,=\,2^{2N+1}\,.
\end{equation}
The finite cardinality of the group $C_D$ ensures that there exists a basis where all matrices $\Gamma_{\mu}$ and their products can be taken unitary. From \cref{eq:gammadagger}, one has in this basis $\Gamma_{\mu}^{\dagger}=\Gamma_{\mu}$ and $\Gamma_{\mu}^{\intercal}=\Gamma_{\mu}^{\ast}$. 

For any finite group, the number of irreducible representations is equal to the number of classes. Thus, the enumeration of classes, $n_c$, of $C_D$ is:
\begin{equation}
\begin{split}
 \{1\},\,\{-1\},\,\{\pm\Gamma_1\},\,\{\pm\Gamma_2\},\,\cdots,\,\{\pm\Gamma_{2N}\},\\
 \,\cdots,\,\{\pm\Gamma_1\Gamma_2\},\,\cdots\,
\{\pm\Gamma_0\}\,,
\end{split}
\end{equation}
which implies that
\begin{equation}
 n_c\,=\,1+\sum_{i=0}^{2N}\binom{2N}{i}\,=\,1+2^{2N}\,.
\end{equation}
We conclude that the finite group $C_D$ has $1+2^{2N}$ irreducible and nonequivalent representations. In order to determine how many unidimensional representations of $C_D$ exist, we recall that in any finite group, the number of unidimensional representations is given by the ratio of number of the elements of the group and the elements of its commutator subgroup. The commutator subgroup of $C_D$ has two elements, namely $[C_D,C_D]\,=\,\{1,-1\}\,$. Thus, the number of unidimensional representations is $\#C_D/\#[C_D,C_D]=2^{2N}$. We then obtain that there is only one non-unidimensional representation for the group $C_D$. The non-trivial representation can be obtained through the relation 
\begin{equation} 
\underbrace{1^2+1^2+\cdots+1^2}_{2^{2N}}\,+\,n^2\,=\,2^{2N+1}\,,
\end{equation}
which implies that $n=2^{N}$. It is clear that all representations of the Clifford algebra are also representations of $C_D$ and therefore the Clifford algebra has only one irreducible representation with dimension $2^{N}$. It is interesting to verify that from a given irreducible representation of the Clifford algebra $\Gamma_{\mu}$, the matrices $\Gamma^{\ast}_{\mu}$, $\Gamma^{\intercal}_{\mu}$, and $-\Gamma^{\intercal}_{\mu}$ form also an irreducible representation of the algebra since
\begin{equation}
\label{eq:gammareps}
\left\{\Gamma^{\ast}_{\mu},\Gamma^{\ast}_{\nu}\right\}\,=\,
\left\{\Gamma^{\intercal}_{\mu},\Gamma^{\intercal}_{\nu}\right\}\,=\,
\left\{-\Gamma^{\intercal}_{\mu},-\Gamma^{\intercal}_{\nu}\right\}\,=\,
2\,\delta_{\mu\nu}\,.
\end{equation}
Due to the fact there is only one irreducible representation of the Clifford algebra, these three representations must be necessarily equivalents. Thus, there must exist matrices $B$ and $C$ such that
\begin{subequations}
 \label{eq:Cdef}
 \begin{align}
 \label{eq:Cdefa}
 B^{-1}\,\Gamma_{\mu}^{\ast}\,B\,&=\,\Gamma_{\mu}\,,\\
 \label{eq:Cdefb}
 B^{-1}\,\Gamma_{\mu}^{\intercal}\,B\,&=\,\Gamma_{\mu}\,,\\
 C^{-1}\,\Gamma_{\mu}^{\intercal}\,C\,&=\,-\Gamma_{\mu}\,.
 \end{align}
 \end{subequations}
Note that the operator $B$ is the same for both \cref{eq:Cdefa,eq:Cdefb} since $\Gamma_{\mu}^{\intercal}=\Gamma_{\mu}^{\ast}$ and the operator $C$ can be written as
\begin{equation}
\label{eq:C}
C=B\Gamma_0\,,
\end{equation}
apart from an overall complex factor. This derivation ensures the existence of the operator $B$, which was used in \cref{eq:needB}. 
\medskip

 We end this appendix by showing that $B$ and $C$ can be taken unitary and they are either antisymmetric or symmetric matrices, depending of the value of $N$. Taking the fact that $(\Gamma_{\mu}^{\intercal})^{\intercal}=\Gamma_{\mu}$ and using~\cref{eq:Cdefa} one concludes 
 \begin{equation}
 \left(B^{-1}B^{\intercal}\right)^{-1} \Gamma_{\mu}\left(B^{-1}B^{\intercal}\right)=\Gamma_{\mu}\,,
 \end{equation}
which implies that $B^{-1}B^{\intercal}$ commutes with all elements of the spinor representation $U(\omega)$ and therefore by the first Schur lemma one has 
\begin{equation}
\label{eq:e}
B^{-1}B^{\intercal}=\epsilon\,\mathbf{1}\,.
\end{equation}
Without the loss of generality, we can refactor the matrix $B$ to have $|B|=1$. This choice implies that $|\epsilon|=1$ and 
\begin{equation}
\label{eq:B}
B^{\intercal}=\epsilon B\,.
\end{equation}
The constant $\epsilon$ is in fact real, since transposing \cref{eq:e} one has
\begin{equation}
\epsilon\,=\,B\,(B^{-1})^{\intercal}\,=\,B\,(B^{\intercal})^{-1}\,=\,\epsilon^{\ast}\,.
\end{equation}
Thus, one has $\epsilon=\pm1$, where the sign will depend only on $N$. Instead, if we use the relation $(\Gamma_{\mu}^{\ast})^{\ast}=\Gamma_{\mu}$, we get
\begin{equation}
 \left[B^{\ast}B,\, \Gamma_{\mu}\right]\,=\,0\,.
 \end{equation}
The product $B^{\ast}B$ then commutes with all elements of the spinor representation $U(\omega)$. The first Schur lemma obliges that 
\begin{equation}
\label{eq:ep}
B^{\ast}B=\epsilon'\,\mathbf{1}\,,
\end{equation}
with $\epsilon'=\pm1$, since $|B|=1$ and taking the conjugation of~\cref{eq:ep}
\begin{equation}
{\epsilon'}^{\ast}\,=\,BB^{\ast}\,=\,BB^{\ast}BB^{-1}\,=\,\epsilon'\,.
\end{equation}
The relation between $\epsilon'$ and $\epsilon$ can be established by observing that $(\Gamma_{\mu}^{\intercal})^{\ast}=(\Gamma_{\mu}^{\ast})^{\intercal}$, which implies
\begin{equation}
 \left[B^{\dagger}B,\, \Gamma_{\mu}\right]\,=\,0\,,
 \end{equation}
The combination $B^{\dagger}B$ commutes with all elements of the spinor representation $U(\omega)$. The first Schur lemma leads to $B^{\dagger}B=\lambda\mathbf{1}$ with $\lambda=1$. This can be obtained since $|B|=1$ (implying $|\lambda|=1$) and for any vector $\ket{v}$ with norm~$1$ one has
\begin{equation}
\lambda\,=\,\bra{v}B^{\dagger}B\ket{v}\,=\,\left|B\ket{v}\right|^2\,\geq0\,.
\end{equation}
We then obtain that $B$ is unitary and as a consequence that $C$ is also unitary. Moreover, using the unitarity of $B$,
\begin{equation}
\mathbf{1}\,=\,B^{\intercal}B^{\ast}\,=\,\epsilon\epsilon'BB^{-1}\,,
\end{equation}
one concludes that $\epsilon'=\epsilon$. 

One may raise the question how to relate the sign of $\epsilon$ with the dimension of $\mathsf{SO}(2N)$. We start by noting that 
\begin{equation}
\Gamma_0^{\intercal}\,=\,B\,\Gamma_0\,B^{-1}=\,C\,\Gamma_0\,C^{-1}\,,
\end{equation}
and one deduces, using \cref{eq:C}, that
\begin{equation}
C^{\intercal}\,=\,\epsilon\, C\,.
\end{equation} 
If one takes into account \cref{eq:invgamma}, one is able to write the following relation
\begin{equation}
\left(C\,\Gamma_{\mu_1}\Gamma_{\mu_2}\cdots\Gamma_{\mu_p}\right)^{\intercal}\,=\,
\epsilon\,(-1)^{\frac{p(p+1)}2}
C\,\Gamma_{\mu_1}\Gamma_{\mu_2}\cdots\Gamma_{\mu_p}\,, 
\end{equation}
for $p\leq2N$. The above relation implies that the matrices $C\,\Gamma_{\mu_1}\Gamma_{\mu_2}\cdots\Gamma_{\mu_p}$ are either symmetric or antisymmetric matrices. On the other hand, the set $\{C\,\Gamma_{\mu_1}\Gamma_{\mu_2}\cdots\Gamma_{\mu_p}\}$ forms a basis of the vector space of all $2^N\times2^N$ complex matrices, since all order product of distinct $\Gamma$-matrices are $2^N$ linearly independent matrices. Thus, the number of independent antisymmetric matrices is given by
\begin{equation}
\sum_{p=0}^{2N}\frac12\left[1-\epsilon\,(-1)^{\frac{p(p+1)}2}\right]\binom{2N}{p}\,.
\end{equation}
For any $2^N\times2^N$ complex matrices, the number of independent antisymmetric matrices is just given by $2^{N}(2^N-1)/2$, therefore one has
\begin{equation}
\sum_{p=0}^{2N}\frac12\left[1-\epsilon\,(-1)^{\frac{p(p+1)}2}\right]\binom{2N}{p}\,=\,\frac{2^{N}\bigl(2^N-1\bigr)}2\,.
\end{equation}
The above equation gives a closed relation between $\epsilon$ and $N$. After some algebraic simplifications, one gets\begin{equation}
\label{eq:espilonN}
\epsilon\,=\,\frac{1}{2^{N+1}}\sum_{p=0}^{2N}(-1)^{\frac{(p-2)(p-1)}2}\binom{2N}{p}\,.
\end{equation}
Taking into account the following relation
\begin{equation}
(-1)^{\frac{(p-2)(p-1)}2}\,=\,-\frac12\biggl[(1+i)\,i^p\,+\,(1-i)\,(-i)^p\biggr]\,,
\end{equation}
that can be verified by mathematical induction, one derives
\begin{equation}
\label{eq:espilonN}
\epsilon\,=\,\sqrt{2}\cos\frac{\pi}4(2N+1)\,.
\end{equation}
This equation has the period $N\rightarrow N+4$, and implies that $\epsilon=1$ for $N=3,\,4\pmod4$, while $\epsilon=-1$ for $N=1,\,2\pmod4$. For instance, in the case of $\mathsf{SO}(10)$, one has $B^{\intercal}=-B$ and $C^{\intercal}=-C$.

%%%%%%%%%%%%%%%%%%%%%%%%%%%%%%%%%%%%%%
\section{Clifford vs. Grassmann algebrae}
 \label{sec:CliffvsGrass}
 
In this appendix, we show the existence of a one-to-one correspondence between Clifford and Grassmann algebrae, apart from an overall complex phases. In order to demonstrate this result, let us take an arbitrary $\Gamma$-matrix pair $\Gamma_{\mu}$ and $\Gamma_{\nu}$ of an abritrary Clifford algebra given in \cref{eq:ca}. Without the loss of generality, we shall take two consecutive gamma matrices,
\begin{equation}
\left\{\Gamma_{2j-1},\Gamma_{2j}\right\}=0\,,\quad \Gamma_{2j-1}^2=\eta_{2j-1}\mathbf{1}\,,\quad \Gamma_{2j}^2=\eta_{2j}\mathbf{1}\,.
\end{equation}
We identify then two possibilities $\eta_{2j-1}=\eta_{2j}$ and $\eta_{2j-1}=-\eta_{2j}$. It is then straightforward to verify that there exist a linear combination $b_j=\alpha\,\Gamma_{2j-1}+\beta\,\Gamma_{2j}$ such that they generate a Grassmann algebra,
\begin{equation}
\left\{b_j,b_j^{\dagger}\right\}=\mathbf{1}\,,\quad b_j^2\,=\,0\,.
\end{equation}
Thus, when $\eta\equiv\eta_{2j-1}=\eta_{2j}$ one has
\begin{equation}
\label{eq:Gb}
b_j\,=\,\frac12\left(i\,\Gamma_{2j-1} \,+\,\Gamma_{2j}\right)\,,
\end{equation}
which then implies
\begin{equation}
\label{eq:Gbt}
b_j^{\dagger}\,=\,\frac{\eta}2\left(-i\,\Gamma_{2j-1} \,+\,\Gamma_{2j}\right)\,,
\end{equation}
where we have used the result that $\Gamma^{\dagger}_a=\eta_a\Gamma_a\,$. The two \cref{eq:Gb,eq:Gbt} can be easily inverted so that the $\Gamma_{2j-1}$ and $\Gamma_{2j}$ are written in terms of $b_j$ and $b_j^{\dagger}$ as
\begin{equation}
\Gamma_{2j-1}\,=\,-i(b_j\,-\,\eta\,b^{\dagger}_j)\,,\quad
\Gamma_{2j}\,=\,b_j\,+\,\eta\,b^{\dagger}_j\,.
\end{equation}
In the case of having $\eta_{2j-1}=-\,\eta_{2j}$, one has instead
\begin{equation}
b_j\,=\,\frac12\left(\Gamma_{2j-1} \,+\,\Gamma_{2j} \right)\,,\quad
b_j^{\dagger}\,=\,\frac{\eta}2\left(\Gamma_{2j-1} \,-\,\Gamma_{2j} \right)\,,
\end{equation}
or the inverted system of equations
\begin{equation}
\Gamma_{2j-1}\,=\,b_j\,-\,\eta\,b^{\dagger}_j\,,\quad
\Gamma_{2j}\,=\,b_j\,+\,\eta\,b^{\dagger}_j\,.
\end{equation}
The fact that two distinct pair of $\Gamma$-matrices anticommutes, it guarantees that the operators $b_j$ and $b^{\dagger}_j$, thus constructed, satisfy fully the Grasmann algebra, i.e.,
\begin{equation}
\left\{b_j,b_k^{\dagger}\right\}=\delta_{jk}\,\mathbf{1}\,,\quad 
\left\{b_j,b_k\right\}\,=\,\left\{b^{\dagger}_j,b_k^{\dagger}\right\}\,=\,0\,.
\end{equation}
We have shown that for any Clifford algebra one can write a set of creation and annihilation operators and therefore the \SOSpin library can be easily adapt for those cases.

%%%%%%%%%%%%%%%%%%%%%%%%%%%%%%%%%%%%%%
\section{$\mathsf{SO}(10)$ compendium}
\label{sec:appSO10}

As a matter of completeness, in this section we compile some $\mathsf{SO}(10)$ details and functions not written in the previous chapters. In particular we summarise~\cite{Nath:2001uw,Nath:2001yj} the action of $\Gamma$-matrices on the Higgs fields, $\phi$, as well as the action of the operator $B$ on $\bra{\psi^\ast_+}$ and $\bra{\psi^\ast_-}$\,.

The generic expression for the operator $B$ is given in \cref{eq:Bopgen} and in what follows we list the action of the operator $B$ on $\bra{\psi^\ast_+}$ and $\bra{\psi^\ast_-}$ given in \cref{eq:psi+,eq:psi-,eq:plusconj,eq:bram}, respectively.
\begin{equation}
\label{eq:plusconjB}
\begin{aligned}
\Bra{\Psi^{\ast}_{+}}B=&-i\,\overline{\psi}_{n}\Bra{0}b_n
-\frac{i}{12}\varepsilon^{njklm}\psi_{nj}\Bra{0}b_k b_l b_m \\
&\,-i\,\psi\Bra{0}b_1 b_2 b_3 b_4 b_5\,,
\end{aligned}
\end{equation}
and
\begin{equation}
\begin{aligned}
\Bra{\Psi^{\ast}_{-}}B=&\,\frac{i}{24}\varepsilon^{njklm}\psi^{n}\Bra{0}b_j b_k b_l b_m
\,+\,\frac{i}{2}\overline{\psi}_{nj}\Bra{0}b_n b_j
+i\,\overline{\psi}\Bra{0}\,.
\end{aligned}
\end{equation}
One sees from the equations above that the action of $B$ on the bras $\bra{\psi^\ast_+}$ and $\bra{\psi^\ast_-}$ gives expressions which are consistent with the $\mathsf{SU}(5)$ convention for upper and lower indices.

We compile below the action of $\Gamma$-matrices on Higgs representations of different dimensions, which follows from \cref{eq:basictheo} and Refs.~\cite{Nath:2001uw,Nath:2001yj}. For the $\mathsf{10}$-dim Higgs representation, $\phi_\mu$, one has,
\begin{equation}
\Gamma_{\mu}\phi_{\mu}=b^{\dagger}_n\,\phi_{c_n}+b_n\,\phi_{\bar{c}_n},
\end{equation}
with the following normalisation coming from the $\mathsf{SU}(5)$ notation
\begin{equation}
\phi_{c_n}=\sqrt{2}H^n,\quad \phi_{\bar{c}_n}=\sqrt{2}H_n.
\end{equation}

For the $\mathsf{45}$-dim representation~\cite{Nath:2001yj}, $\phi_{\mu\nu}$, one has,
\begin{equation}
\label{eq:gen45}
i\,\Sigma_{\mu \nu}\phi_{\mu \nu}=
b_i b_j \phi_{\bar{c}_i\bar{c}_j} + b^{\dagger}_ib^{\dagger}_j\phi_{c_i c_j}
+2b^{\dagger}_ib_j\phi_{c_i\bar{c}_j}-\phi_{c_n\bar{c}_n}
\,,
\end{equation}
where 
\begin{equation}
\begin{array}{l@{\quad}l}
\phi_{c_n\bar{c}_n}=h\,, & \phi_{c_i\bar{c}_j}=h^i_j + \frac{1}{5}\delta^i_j h\,, \\
\phi_{c_i c_j}=h^{ij}\,, & \phi_{\bar{c}_i\bar{c}_j}=h_{ij}\,, 
\end{array}
\end{equation}
with the normalisation
\begin{equation}
\begin{array}{l@{\quad}l}
h=\sqrt{10}\,H\,, &
h^{ij}=\sqrt{2}\,H^{ij},\\
h_{ij}=\sqrt{2}\,H_{ij}\,, &
h^i_j=\sqrt{2}\,H^i_j\,.
\end{array}
\end{equation}

For the $\mathsf{120}$-dim Higgs field~\cite{Nath:2001yj}, $\phi_{\mu\nu\lambda}$, one has,
\begin{equation}
\label{eq:gen120new}
\begin{aligned}
\Gamma_{\mu}\Gamma_{\nu}&\Gamma_{\lambda}\phi_{\mu \nu \lambda}=b_i b_j 
b_k \phi_{\bar{c}_i\bar{c}_j\bar{c}_k}+b^{\dagger}_ib^{\dagger}_jb^{\dagger}_k\phi_{c_ic_jc_k}\\
&+3(b^{\dagger}_ib_j b_k\phi_{c_i\bar{c}_j\bar{c}_k} + b^{\dagger}_i b^{\dagger}_jb_k\phi_{c_i c_j\bar{c}_k})\\
&+(3b_i\phi_{\bar{c}_n c_n\bar{c}_i} + 3b^{\dagger}_i\phi_{\bar{c}_n c_n\bar{c}_i})\,,
\end{aligned}
\end{equation}
where
\begin{align}
\phi_{c_i c_j\bar{c}_k}&=h^{ij}_k+\frac{1}{4}\left(\delta^i_k h^j-\delta^j_k h^i\right)\,,\\
\label{eq:c12}
\phi_{c_i\bar{c}_j\bar{c}_k}&= h_{jk}^i-\frac{1}{4}\left(\delta^i_j h_k-\delta^i_k h_j\right)\,,\\
\phi_{c_ic_jc_k}&= \varepsilon^{ijklm}h_{lm}\,,\\
\phi_{\bar{c}_i\bar{c}_j\bar{c}_k}&=\varepsilon_{ijklm}h^{lm}\,,\\
\phi_{\bar{c}_n c_n\bar{c}_i}&=h^i\,, \\
\label{eq:c16}
\phi_{\bar{c}_n c_n\bar{c}_i}&=-h_i\,,
\end{align}
with the following normalisation
\begin{equation}
\begin{array}{l@{\quad}l}
h^i&=\frac{4}{\sqrt{3}}\,H^i\,,\quad h_i=\frac{4}{\sqrt{3}}\,H_i\,,\\
h^{ij}&=\frac{1}{\sqrt{3}}\,H^{ij}\,,\quad h_{ij}=\frac{1}{\sqrt{3}}\,H_{ij}\,,\\
h^{ij}_k&=\frac{2}{\sqrt{3}}\,H^{ij}_k\,,\quad h^i_{jk}=\frac{2}{\sqrt{3}}\,H^i_{jk}\,.
\end{array}
\end{equation}
Note that we have corrected the signs of the tensors in~\cref{eq:c12,eq:c16}.

For the 210-dim representation~\cite{Nath:2001yj}, $\phi_{\mu\nu\rho\lambda}$, one has,
\begin{equation}
\label{eq:gen210}
\begin{aligned}
\Gamma_{\mu}\Gamma_{\nu}\Gamma_{\rho}\Gamma_{\lambda}
\,\phi_{\mu \nu \rho \lambda}=&4\,b^{\dagger}_ib^{\dagger}_jb^{\dagger}_kb_l\phi_{c_ic_jc_k\bar{c}_l}
+4\,b^{\dagger}_ib_jb_kb_l\phi_{c_i\bar{c}_j\bar{c}_k\bar{c}_l}\\
&+b^{\dagger}_ib^{\dagger}_jb^{\dagger}_kb^{\dagger}_l\phi_{c_ic_jc_kc_l}
+b_i b_j b_k b_l\phi_{\bar{c}_i\bar{c}_j\bar{c}_k\bar{c}_l}\\
&-6\,b^{\dagger}_ib^{\dagger}_j\phi_{c_i c_jc_m\bar{c}_m}
+6\,b_ib_j\phi_{\bar{c}_i\bar{c}_j\bar{c}_m c_m}\\
&+3\,\phi_{c_m\bar{c}_m c_n\bar{c}_n} 
-12\,b^{\dagger}_ib_j\phi_{c_i\bar{c}_jc_m\bar{c}_m}\\
&+6\,b^{\dagger}_ib^{\dagger}_jb_kb_l\phi_{c_ic_j\bar{c}_k\bar{c}_l}\,,
\end{aligned}
\end{equation}
where
\begin{equation}
\begin{aligned}
&\begin{aligned}
\phi_{c_m\bar{c}_m c_n\bar{c}_n}&= h, & \phi_{c_i\bar{c}_jc_m\bar{c}_m}&=h^i_j+ \frac{1}{5}\delta^i_j h,\\
\phi_{\bar{c}_i\bar{c}_j\bar{c}_k\bar{c}_l}&= \frac{1}{24}\varepsilon_{ijklm}h^m, & \phi_{c_ic_jc_kc_l}&= \frac{1}{24}\varepsilon^{ijklm}h_m,\\
\phi_{c_i c_jc_m\bar{c}_m}&= h^{ij}, & \phi_{\bar{c}_i\bar{c}_j\bar{c}_m c_m}&= h_{ij},
\end{aligned}\\
&\begin{aligned}
\phi_{c_ic_j\bar{c}_k\bar{c}_l}\,=\,& h^{ij}_{kl}+\frac{1}{3}\left(\delta^i_l h^j_k -\delta^i_k h^j_l +\delta^j_k h^i_l - \delta^j_l h^i_k \right) \\
&+ \frac{1}{20}\left(\delta^i_l \delta^j_k - \delta^i_k \delta^j_l\right)h,\\
\phi_{c_ic_jc_k\bar{c}_l} =&h^{ijk}_l+\frac{1}{3}\left(\delta^k_l h^{ij}-\delta^j_l h^{ik} +\delta^i_l h^{jk}\right) ,
\end{aligned}\\
&\phi_{c_i\bar{c}_j\bar{c}_k\bar{c}_l}=h^i_{jkl}+\frac{1}{3}\left(\delta_i^l h_{jk}-\delta^i_k h_{jl} +\delta^i_j h_{jk}\right)\,,
\end{aligned}
\end{equation}
with the normalisation
\begin{equation}
\begin{array}{l@{\quad}l@{\quad}l}
h= \frac{4\sqrt{5}}{\sqrt{3}}H\,, & h^i=8\sqrt{6}H^i\,, & h_i=8\sqrt{6}H_i\,,\\[2mm]
h^{ij}= \sqrt{2}H^{ij}\,, & h_{ij}= \sqrt{2}H_{ij}\,, & h^i_j= \sqrt{2}H^i_j\,,\\[2mm]
h^{ijk}_l=\frac{\sqrt{2}}{\sqrt{3}} H^{ijk}_l\,, & h^i_{jkl}= \frac{\sqrt{2}}{\sqrt{3}}H^i_{jkl}\,,
& h^{ij}_{kl}=\frac{\sqrt{2}}{\sqrt{3}} H^{ij}_{kl}\,.
\end{array}
\end{equation}

Finally, for the irreducible representations $\mathsf{126}$ ($\phi_{\mu\,\nu\,\lambda\,\rho\,\sigma}$) and $\mathsf{\overline{126}}$ ($\overline{\phi}_{\mu\,\nu\,\lambda\,\rho\,\sigma}$), as it was observed in~\cref{subsec:Yukawa}, one needs only to take into account the reducible $\mathsf{252}$ representation, since only one of its irreducible components survives~\cite{Nath:2001uw},
\begin{equation}
\begin{aligned}
\Gamma_{\mu} &\Gamma_{\nu} \Gamma_{\lambda} \Gamma_{\rho} \Gamma_{\sigma}\, \Delta_{\mu\,\nu\,\lambda\,\rho\,\sigma}=\varepsilon_{ijklm} b_i b_j b_k b_lb_m h\\
&+\varepsilon^{ijklm}\, b_i^{\dagger} b_j^{\dagger}b_k^{\dagger}
b_l^{\dagger}b_m^{\dagger} \bar h +15\, b_i^{\dagger} h^i\\
&-20\, b_i^{\dagger} b_n^{\dagger}b_n h^i 
+5 b_i^{\dagger} b_n^{\dagger}b_n b_m^{\dagger}b_m h^i\\
&+ 15\, b_i h_i
-20\, b_n^{\dagger}b_n b_i h_i 
+5\, b_n^{\dagger}b_n b_m^{\dagger}b_m b_i h_i)\\
&+10\, (b_i^{\dagger} b_j^{\dagger}b_k^{\dagger}h^{ijk}
-b_i^{\dagger} b_j^{\dagger}b_k^{\dagger}b_n^{\dagger}b_n h^{ijk}\\
& +b_ib_jb_k h_{ijk} - b_n^{\dagger}b_n b_ib_jb_k h_{ijk})\\
& + (60\, b_i^{\dagger} b_j^{\dagger}b_k h^{ij}_k
-30\, b_i^{\dagger} b_j^{\dagger}b_k b_n^{\dagger}b_n h^{ij}_k\\
&+ 60\, b_i^{\dagger} b_j b_k h^{i}_{jk}
-30 b_n^{\dagger}b_n b_i^{\dagger} b_j b_k h^{i}_{jk})\\
& + (5\,b_i^{\dagger} b_j^{\dagger}b_k^{\dagger}
b_l^{\dagger}b_m h^{ijkl}_m
+ 5\,b_i^{\dagger} b_jb_kb_lb_m h^{i}_{jklm})\\
&+(10\,b_i^{\dagger} b_j^{\dagger}b_k^{\dagger}
b_lb_m h^{ijk}_{lm}
+ 10\,b_i^{\dagger} b_j^{\dagger}b_kb_lb_m h^{ij}_{klm})\,,
\label{eq:gamma}
\end{aligned}
\end{equation}
where 
\begin{equation}
\begin{aligned}
&h=\frac{2}{\sqrt{15}}\,H\,,\quad h^i=\frac{4\sqrt{2}}{\sqrt{5}}H^i\,,\\ 
&h^{ijk}=\frac{\sqrt{2}}{\sqrt{15}}\varepsilon^{ijklm}\,H_{lm}\,,\quad
h^i_{jklm}=\frac{\sqrt{2}}{\sqrt{15}}\varepsilon_{jklmn}\,H^{ni}_{(S)}\,,\\
&h^i_{jk}=\frac{2\sqrt{2}}{\sqrt{15}}\,H^i_{jk}\,,\quad
f^{ijk}_{lm}=\frac{2}{\sqrt{15}}\,H^{ijk}_{lm}\,.
\end{aligned}
\end{equation}
The field $H^{ij}_{(S)}$ denotes the $\mathsf{15}$ representation of $\mathsf{SU}(5)$, which is a symmetric tensor.

%%%%%%%%%%%%%%%%%%%%%%%%%%%%%%%%%%%%%%
\section{Low-level implementations of the \texorpdfstring{\protect\SOSpin}{SOSpin} library}
\label{sec:LOW}

In this section we list and describe all functions in the header files \code{dlist.h}, \code{index.h}, \lstinline[basicstyle=\normalsize\ttfamily,emphstyle=]!braket.h!, \code{form.h} and \code{son.h}.

%%%%%%%%%%%%%%%%%%%%%%%%%%%%%%%%%%%%%%
\paragraph{\code{dlist.h}}
\label{subsec:dlist}

\begin{itemize}
\item \code{DList::DList(void)}\\
\code{DList::DList(const DList \&L)}\\
\code{DList::DList(int type, int i, int j)}\\
\code{DList::DList(int type, int i)}\\
\code{DList::~DList(void)}\\[2mm]
The prototypes stand for the default, copy and specific constructors and the destructor, respectively, within the class \code{DList}. Note that the argument \code{type} defines the type of the node element. The arguments \code{i} and \code{j} are at the most two indices to complete the element. Pointers to the \code{noList}-type fields are initialised with \code{NULL}.

\item \code{void DList::clear()}\\[2mm]
Deletes all nodes from \code{DList}.

\item \code{void DList::negate()}\\[2mm]
Changes the sign of \code{DList}.

\item \code{void DList::add(elemType)}\\[2mm]
Adds one node at the end of \code{DList} (scans the list) and updates the actual pointer to be the last node.

\item \code{void DList::add\_begin(elemType)}\\[2mm]
Adds one node in the beginning of \code{DList} and updates the actual pointer to be the new first node.

\item \code{void DList::add\_begin(elemType)}\\[2mm]
Adds one node in the beginning of \code{DList} and updates the actual pointer to be the new first node.

\item \code{void DList::add\_end(elemType)}\\[2mm]
Adds one node in the end of \code{DList} and updates the actual pointer to be the last node.

\item \code{void DList::set(elemType)}\\[2mm]
Sets data (\code{elemType}) of the node being pointed by actual pointer.

\item \code{void DList::set\_begin()}\\[2mm]
Changes actual pointer to point at the first element of \code{DList} (beg pointer).

\item \code{void DList::set\_end()}\\[2mm]
Changes actual pointer to point at the last element of \code{DList} (end pointer).

\item \code{void DList::set\_sign(int i)}\\[2mm]
Sets the sign of \code{DList}.

\item \code{void DList::join(DList\& L)}\\[2mm]
Joins a \code{DList} to the end of the current \code{DList} (this) and updates the $actual$ pointer to be the end of the final \code{DList}.

\item \code{void DList::loop\_right()}\\[2mm]
Shifts actual pointer to next node. If actual node is the end node, shift to beg node.

\item \code{void DList::loop\_left()}\\[2mm]
Shifts actual pointer to previous node. If actual node is the beg node, shift to end node.

\item \code{DList DList::rearrange()}\\[2mm]
Creates and returns a new \code{DList} by copying nodes in \code{DList} ordered by type. The nodes that first appear in the new ordered \code{DList} are $\delta$'s (\code{type}=2) and then all other elements: $b$ (\code{type}=0) and $b^\dagger$ (\code{type}=1) unordered. Constant elements are removed.

\item \code{void DList::remove(unsigned int type)}\\[2mm]
Removes the first element with \code{data.get\_type()==type} found in \code{DList}. Updates actual pointer to be the first node.

\item \code{void DList::remove\_actual()}\\[2mm]
Removes the element for which the actual pointer, $actual$, is pointing at in \code{DList}.

\item \code{void DList::shift\_right()}\\[2mm]
Shifts actual pointer to next node. If actual node is the end node, stops.

\item \code{void DList::shift\_left()}\\[2mm]
Shifts actual pointer to previous node. If actual node if first node ($begin$), stops.

\item \code{void DList::swap\_next()}\\[2mm]
Swaps the actual node with the next node of \code{DList}.

\item \code{elemType DList::get()}\\[2mm]
Returns elemtype of the node being pointed by actual (current element).

\item \code{int DList::getSign()}\\[2mm]
Returns the sign of \code{DList}.

\item \code{vector<int> DList::getIds()}\\[2mm]
Creates and returns an integer vector sequence container with the ids (data fields) of $b$'s and $b^\dagger$'s elements.

\item \code{void DList::getBandBdaggerIds(bool \&BandBd, vector <string> \&id0, vector<string> \&id1, int \&sign)}\\[2mm]
Updates integer vector sequence containers \code{id0} and \code{id1} with ids (data fields) of $b$'s and $b^\dagger$'s elements, respectively; the parameter \code{sign} is updated with the sign of \code{DList} and \code{BandBd} is a boolean which is true if \code{DList} contains at least one $b$ or $b^\dagger$ and false otherwise.

\item \code{void DList::getDeltaIds(bool \&AllDeltas, vector<string> \&id0, vector<string> \&id1, int \&sign)}\\[2mm]
Updates integer vector sequence containers \code{id0} and \code{id1} with the first and second ids (data fields) of $\delta$ elements, respectively; the parameter \code{sign} is updated with the sign of \code{DList} and \code{AllDeltas} is a boolean which is true if all elements in \code{DList} are of $\delta$ type and false otherwise.

\item \code{void DList::getBandBdaggerAndDeltasIds( vector <string> \&id0, vector<string> \&id1, vector<string> \&id2, vector<string> \&id3, int \&sign)}\\[2mm]
Updates integer vector sequence containers \code{id0} and \code{id1} with ids (data fields) of $b$'s and $b^\dagger$'s elements, respectively; updates integer vector sequence containers \code{id2} and \code{id3} with first and second data fields of $\delta$'s elements, respectively. The parameter \code{sign} is updated with the sign of \code{DList}.

\item \code{int DList::numDeltas()}\\[2mm]
Returns the number of elements of type $\delta$ (\code{type}=2).

\item \code{int DList::numBs()}\\[2mm]
Returns the number of elements of type $b$ (\code{type}=0).

\item \code{bool DList::search\_last(unsigned int type1)}\\[2mm]
Search the last element with \code{data.get\_type()==type1} found in \code{DList}. 
Returns true a node was found.

\item \code{bool DList::search\_first(unsigned int type1)}\\[2mm]
Search the first element with \code{data.get\_type()==type} found in \code{DList} and returns true if the symbol is not the first element, and false otherwise.

\item \code{bool DList::search\_first(unsigned int type0, unsigned int type1)}\\[2mm]
Checks for the order of appearance in \code{DList} of types \code{type0} and \code{type1}. Returns true if order of appearance is the same as the parameter's order and false otherwise.

\item \code{bool DList::search\_elem(unsigned int type1)}\\[2mm]
Searches for the element with \code{data.get\_type()==type} found in \code{DList}. Returns true if the element is found, false otherwise.

\item \code{bool DList::check()}\\[2mm]
Verifies if the number of $b$'s and $b^\dagger$'s matches and whether the number each one is less or equal than $N$ of $\mathsf{SO}(2N)$.

\item \code{bool DList::checkDeltaIndex()}\\[2mm]
Checks the indexes of $\delta$ elements and if the indexes in $\delta$ are equal. Returns true if each $\delta$ is not zero, false otherwise.

\item \code{bool DList::check\_num()}\\[2mm]
Verifies if the number of $b$'s and $b^\dagger$'s is less or equal than $N$ of $\mathsf{SO}(2N)$ if so the function returns true if not returns false.

\item \code{bool DList::check\_same\_num()}\\[2mm]
Verifies if the number of $b$'s and $b^\dagger$'s matchs; returns true if they match and false otherwise.

\item \code{bool DList::isActualLast()}\\[2mm]
Returns true if actual pointer is pointing to the last ($end$) node of \code{DList}.

\item \code{bool DList::isEmpty()}\\[2mm]
Returns true if \code{DList} has no nodes.

\item \code{bool DList::hasNoDeltas()}\\[2mm]
Returns true if there is no elements of type $\delta$ in \code{DList}.

\item \code{bool DList::hasOnlyDeltas()}\\[2mm]
Returns true if all nodes in \code{DList} are of $\delta$ type.

\item \code{bool DList::hasRepeatedIndex()}\\[2mm]
Returns true if there is elements with the same id (data fields) in the \code{DList} (repeated ids).

\item \code{DList\& DList::operator=(const DList\& L)}\\[2mm]
Copies a \code{DList}.

\item \code{const DList DList::operator-()const}\\[2mm]
Negates operator, change sign of \code{DList}.

\item \code{friend DList* copy(DList *L)}\\[2mm]
Creates and returns a pointer to a new copy of a \code{DList}.

\item \code{friend DList contract\_deltas(DList \&L, bool braketmode)}\\[2mm]
Applies the following identity $b_i * b^\dagger_j = \delta_{i,j} - b^\dagger_j * b_i$, input \code{DList\&L} keeps delta term and the function returns the swapped term; the parameter \code{braketmode}- if true and if last element in \code{DList} is a $b$, then the \code{L} is cleared. Returns expression with $b_i * b^\dagger_j$ swapped or empty expression if $b$ is the last term in \code{L}.

\item \code{friend DList ordering(DList \&L, bool braketmode)}\\[2mm]
Order only the $b$'s (to the left hand side) and $b^\dagger$'s (to the right hand side) terms.
     Applies the following identity $b^\dagger_j * b_i = \delta_{i,j} - b_i * b^\dagger_j$, input \code{DList} $L$ keeps delta term and function returns the swapped term. \code{braketmode} - if true and if last element in \code{DList} is a $b$, then \code{L} is cleared. Returns expression with $b^\dagger_j * b_i$ swapped or empty expression if $b$ is the last term in \code{L}.

\item \code{friend string printDeltas(DList \&L)}\\[2mm]
Creates and returns a string with the deltas and constants of a \code{DList}.

\item \code{friend DList\& operator*(DList\& L, elemType j)}\\[2mm]
Adds element \code{j} to the end of \code{DList} and returns a pointer to \code{DList}.

\item \code{friend DList operator*(const DList\& L, const DList \&M)}\\[2mm]
Creates and returns a new \code{DList} that joins two \code{DLists} by order of parameters.

\item \code{friend DList\& operator-(DList\& L, elemType j)}\\[2mm]
Negates the sign of \code{DList} and adds the \code{elemtype j} at end of it.

\item \code{friend DList\& operator,(DList\& L, elemType j)}\\[2mm]
Adds element \code{j} to the end of \code{DList} and returns a pointer to \code{DList}.

\item \code{friend ostream\& operator<<(ostream\& out, DList \&L)}\\[2mm]
Sends to output \code{stream ostream} a string with the corresponding expression of the \code{DList}.

\item \code{friend ostream\& operator<<(ostream\& out, DList *L)}\\[2mm]
Sends to output \code{stream ostream} a string with the corresponding expression of the \code{DList}.

\item \code{friend DList\& operator<<(DList \&L, elemType j)}\\[2mm]
Adds element \code{j} to the end of \code{DList} and returns a pointer to \code{DList}.

\item \code{friend DList\& operator<<(DList \&L, DList \&M)}\\[2mm]
Copies \code{DList}; creates and returns a new \code{DList} with nodes of both old and new \code{DLists}. Sign is the product of both products.

\item \code{friend bool operator==(DList \&L, DList \&M)}\\[2mm]
Returns true if two \code{DLists} are equal.
\end{itemize}

%%%%%%%%%%%%%%%%%%%%%%%%%%%%%%%%%%%%%%
\paragraph{\code{index.h}}

\begin{itemize}
\item \code{int newIdx(int i)}\\[2mm]
See definition in \cref{sec:generalfunctions}.

\item \code{int newIdx(string i)}\\[2mm]
Stores a new index of type \code{string}.

\item \code{void newId(string i)}\\[2mm]
Stores a new index of type \code{string}.

\item \code{string getIdx(int i)}\\[2mm]
Returns the index placed at the position \code{i}.

\item \code{int Idx\_size()}\\[2mm]
Returns index list size.

\item \code{string IndexList()}\\[2mm]
Returns index list in string of the form "Indices ?,...,?".

\item \code{void printIDX()}\\[2mm]
Prints index list.

\item \code{string makeId(string a, int id)}\\[2mm]
Returns \code{a}+\code{id} in a string format.

\item \code{template<class T> string ToString(T number)}\\[2mm]
Converts to string.
\end{itemize}

%%%%%%%%%%%%%%%%%%%%%%%%%%%%%%%%%%%%%%
\paragraph{\code{braket.h}}

\begin{itemize}
\item \code{void setSimplifyIndexSum()} \\[2mm]
See definition in \cref{sec:generalfunctions}.

\item \code{void unsetSimplifyIndexSum()} \\[2mm]
See definition in \cref{sec:generalfunctions}.

\item \code{BraketOneTerm::BraketOneTerm()}\\[2mm]
Constructor.

\item \code{BraketOneTerm::BraketOneTerm(const DList &d)}\\[2mm]
Constructor without constant part and index zero; \code{d} is a \code{DList} expression.

\item \code{BraketOneTerm::BraketOneTerm(int indexin, string constpartin, const DList &d)}\\[2mm]
Constructor; \code{indexin} is the index of the expression, 
\code{constpartin} is the constant part and \code{d} is a \code{DList} expression.
		
\item \code{BraketOneTerm::BraketOneTerm(int indexin, string constpartin, list<DList> termin)}\\[2mm]
Constructor; \code{indexin} is the index of the expression, \code{constpartin} is the constant part and \code{termin} is a \code{list<DList>} expression.
									
\item \code{BraketOneTerm::BraketOneTerm(int indexin, string constpartin, BraketOneTerm& termin)}\\[2mm]
Constructor; \code{indexin} is the index of the expression, \code{constpartin} is the constant part and \code{termin} is a \code{BraketOneTerm} expression.
	
\item \code{BraketOneTerm::~BraketOneTerm()}\\[2mm]
Destructor; clears all allocated memory.

\item \code{void BraketOneTerm::clear()}\\[2mm]
Clears all allocated memory and sets default parameters.
		
\item \code{list<DList>& BraketOneTerm::GetTerm()}\\[2mm]
Returns and sets the term part.
 
\item \code{string& BraketOneTerm::GetConst()}\\[2mm]
Returns (and sets) the constant part.
			
\item \code{int& BraketOneTerm::GetIndex()}\\[2mm]
Returns (and sets) the index sum part.
	
\item \code{bool BraketOneTerm::Simplify(OPMode oper)}\\[2mm]
Simplifies current expression term.

\code{oper} term mode (bra, braket, ket or none)

return true if expression term is empty, false otherwise

\item \code{bool BraketOneTerm::checkindex()}\\[2mm]
Checks global index in expression term; returns true if |index| is equal to 0 or $N$ of $\mathsf{SO}(2N)$, otherwise returns false.		
			
\item \code{void BraketOneTerm::expfromForm(string a)}\\[2mm]	
To pass an expression from form.
			
\item \code{void BraketOneTerm::rearrange()}\\[2mm]
Orders nodes of \code{DList} in \code{Braket}. First deltas and then $b$'s and $b^\dagger$'s, and removes the identity node when $\delta$, $b$ or $b^\dagger$ are present.

\item \code{bool BraketOneTerm::isempty()}\\[2mm]
Returns true if expression is \code{empty}.

\item \code{bool BraketOneTerm::EvaluateToDeltas(OPMode oper)}\\[2mm]
Evaluates the expression to deltas; the mode of \code{oper} can be \code{bra}, \code{braket}, \code{ket} or \code{none}. This function returns true if the term is empty (or gives zero) otherwise returns false.

\item \code{bool BraketOneTerm::EvaluateToLeviCivita(OPMode oper)}\\[2mm]
Evaluates the expression to Levi-Civita tensors with eventual $\delta$'s; the mode of \code{oper} can be \code{bra}, \code{braket}, \code{ket} or \code{none}. This function returns true if the term is empty (or gives zero) otherwise returns false.

\item \code{void BraketOneTerm::neg()}\\[2mm]
Negates \code{BraketOneTerm}.

\item \code{BraketOneTerm BraketOneTerm::operator*(const string constval)}\\[2mm]
Overloads \code{operator} for \code{BraketOneTerm * constval}.

\item \code{BraketOneTerm BraketOneTerm::operator*=(const string constval)}\\[2mm]
Overloads \code{operator} for \code{BraketOneTerm *= constval}.

\item \code{BraketOneTerm BraketOneTerm::operator*(const BraketOneTerm &L)}\\[2mm] 
Overloads \code{operator} for \code{BraketOneTerm * L}.

\item \code{BraketOneTerm BraketOneTerm::operator*=(const BraketOneTerm &L)}\\[2mm]
Overload \code{operator} for \code{BraketOneTerm *= L}.

\item \code{friend BraketOneTerm operator-(const BraketOneTerm &L)}\\[2mm]
Negates operator.

\item \code{friend ostream& operator<<(ostream& out, const BraketOneTerm &L)}\\[2mm]
Stream operator.

\item \code{Braket::Braket(void)}\\[2mm]
Constructor; the default expression mode is none.

\item \code{Braket::Braket(const DList &d0)}\\[2mm]
Constructor without constant part and index zero; \code{d0} is \code{DList} expression.

\item \code{Braket::Braket(int id, string a, DList d0)}\\[2mm]
Constructor; the default expression mode is none and \code{id} is the index of the expression, \code{a} is the constant part and \code{d0} is the \code{DList} expression.
			
\item \code{Braket::Braket(int id, string a, DList d0, OPMode op)}\\[2mm]
Constructor, default expression mode is none.

\code{id} - index of the expression

\code{a} - constant part

\code{d0} - \code{DList} expression

\code{op} - Braket type, i.e., bra/ket/braket/none
			
\item \code{Braket::Braket(const Braket &L)}\\[2mm]
Constructor.

\code{L} - \code{Braket} expression
	
\item \code{Braket::Braket(int id, string a, const Braket &L, OPMode op)}\\[2mm]
Constructor.

\code{id} - index of the expression

\code{a} - constant part

\code{L} - \code{Braket} expression, ignores current constant part of \code{L}

\code{op} - \code{Braket} type, i.e., bra/ket/braket/none
	
\item \code{Braket::Braket(BraketOneTerm term)}\\[2mm]
Constructor, default expression mode is none.

\code{term} - \code{BraketOneTerm} expression
		
\item \code{Braket::Braket(BraketOneTerm term, OPMode op)}\\[2mm]
Constructor.

\code{term} - \code{BraketOneTerm} expression

op - \code{Braket} type, i.e., bra/ket/braket/none
	
\item \code{Braket::~Braket()}\\[2mm]
Destructor, clears all allocated memory.

\item \code{void Braket::clear()}\\[2mm]
Clears all allocated memory and sets default parameters.
		
\item \code{void Braket::expfromForm(vector<string> a)}\\[2mm]
To pass an expression from form.
		
\item \code{OPMode& Braket::Type()}\\[2mm]
Returns the current expression type, it also allows to set new expression type. Expression types: bra/ket/braket or none.
		
\item \code{void Braket::mode()}\\[2mm]
Prints the \code{Braket} expression mode, ie, the type of \code{Braket}: bra/ket/braket or none.
	
\item \code{void Braket::evaluate(bool onlydeltas=true)}\\[2mm]
See \cref{sec:generalfunctions}.

\item \code{void Braket::simplify()}\\[2mm]
Simplifies expression. Applies the following rules: $b_i \ket{0} = 0 $ and $\bra{0} b^\dagger_j = 0$. In $\bra{0}... \ket{0}$ the number of $b_i$ must be equal to the number of $b^\dagger_j$. It also checks for numeric deltas and evaluates them.
		
\item \code{void Braket::rearrange()}\\[2mm]
Orders nodes of \code{DList} in Braket. First deltas and then $b$'s and $b^\dagger$'s and removes the identity node when deltas, $b$ or $b^\dagger$ are present.
	
\item \code{void Braket::checkDeltaIndex()}\\[2mm]
Checks index in the deltas.
		
\item \code{void Braket::gindexsetnull()}\\[2mm]
Sets to zero the index sum of each expression term.
		
\item \code{void Braket::checkindex()}\\[2mm]
Checks global index in expression term if \code{setSimplifyIndexSum()} or \code{FlagSimplifyGlobalIndexSum()} is active, returns true if |index| is equal to 0 or $N$ of $\mathsf{SO}(2N)$, otherwise returns false.

\item \code{void Braket::setON()}\\[2mm]
Activates expression term numbering for output writing for each term \code{Local R?=}.
		
\item \code{void Braket::setOFF()}\\[2mm]
Deactivates expression term numbering for output writing for each term \code{Local R?=}.
		
\item \code{int Braket::size()}\\[2mm]
Returns number of terms in current expression.
		
\item \code{BraketOneTerm& Braket::Get(int pos)}\\[2mm]
 Returns expression term at position given by \code{pos}.
		
\item \code{int& Braket::GetIndex(int pos)}\\[2mm]
Returns/sets the index sum of the term given by \code{pos}.
			
\item \code{Braket Braket::operator=(const Braket &L)}\\[2mm]
Overloads operator for \code{Braket = L}.
	
\item \code{Braket Braket::operator+(const Braket &L)}\\[2mm]
Overloads operator for \code{Braket + L}.
	
\item \code{Braket Braket::operator+=(const Braket &L)}\\[2mm]
Overloads operator for \code{Braket += L}.
	
\item \code{Braket Braket::operator-(const Braket &L)}\\[2mm]
Overloads operator for \code{Braket - L}.
		
\item \code{Braket Braket::operator-=(const Braket &L)}\\[2mm]
Overloads operator for \code{Braket -= L}.
	
\item \code{Braket Braket::operator*(const Braket &L)}\\[2mm]
Overloads operator for \code{Braket * L}.
		
\item \code{Braket& Braket::operator*=(const Braket &L)}\\[2mm]
Overloads operator for \code{Braket *= L}.
		
\item \code{Braket Braket::operator*(const string constval)}\\[2mm]
Overloads operator for \code{Braket * constval}, i.e., the constant part.
		
\item \code{Braket Braket::operator*=(const string constval)}\\[2mm]
Overloads operator for \code{Braket *= constval}, i.e., the constant part.
	
\item \code{friend Braket operator-(const Braket &L)}\\[2mm]
Overloads operator for negate, \code{-L}.

\item \codeb{friend OPMode operator*(const OPMode a, const 
OPMode b)}\\[2mm]
Calculates the mode for the multiplication. Returns mode of the multiplication, if this results in an invalid operation the program exit. \code{a} is the mode of the \code{left operand}, \code{b} is the mode of the \code{right operand}.

\item \codeb{friend OPMode operator+(const OPMode a, const OPMode b)}\\[2mm]
Calculates the mode for the sum. Returns mode of the sum, if this results in an invalid operation the program exit. \code{a} is mode of the \code{left operand} e \codeb{b} is mode of the \code{right operand}.

\item \codeb{friend OPMode operator-(const OPMode a, const OPMode b)}\\[2mm]
Calculates the mode for the subtraction. Returns mode of the subtraction, if this results in an invalid operation the program exit. \code{a} is the mode of the \code{left operand} e \codeb{b} is mode of the \code{right operand}. 
			
\item \code{friend ostream& operator<<(ostream& out, const Braket &L)}\\[2mm]
Writes expression to \code{ostream}.
		
\item \code{friend string& operator<<(string& out, const Braket &L)}\\[2mm]
Writes expression to \code{string}.
		
\item \code{friend string& operator+ (string& out, const Braket &L)}\\[2mm]
Writes expression to \code{string}.

\item \code{ostream& operator<<(ostream& out, const OPMode &a)}\\[2mm]
Gets the mode of the expression \code{a} of the current expression and returns the mode in \code{ostream}.
\end{itemize}

%%%%%%%%%%%%%%%%%%%%%%%%%%%%%%%%%%%%%%
\paragraph{\code{form.h}}

\begin{itemize}
\item \code{void setFormRenumber()}\\
\code{void unsetFormRenumber()}\\
\code{void setFormIndexSum()}\\
\code{void unsetFormIndexSum()}\\[2mm]
See \cref{sec:generalfunctions}.

\item \code{string FormField(const string fieldname, const unsigned int numUpperIds, const unsigned int numLowerIds, const FuncProp funcp)}\\[2mm]
See \cref{sec:generalfunctions}.

\item \code{void CallForm(Braket \&exp, bool print=true, bool all=true, string newidlabel="j")}\\[2mm]
See \cref{sec:generalfunctions}.

\item \code{ToForm::ToForm(void)}\\[2mm]
Constructor.
 
\item \code{ToForm::~ToForm()}\\[2mm]
Destructor.
 
\item \code{void ToForm::clear()}\\[2mm]
Clears all allocated memory and sets default values.
 
\item \code{bool ToForm::function(string f)}\\[2mm]
Stores a field name.
 
\item \code{void ToForm::contractions(string f)}\\[2mm]
Stores all field contractions.
 
\item \code{ToForm::string getFC()}\\[2mm]
Returns all type of contractions for the fields.
 
\item \code{string ToForm::getFunction()}\\[2mm]
Returns all the field names.
 
\item \code{void ToForm::setFilename(string name)}\\[2mm]
Sets the beginning of a input/output FORM file.
    
\item \code{string ToForm::file()}\\[2mm]
Returns the beginning of a input/output FORM file.
 
\item \code{string& ToForm::rpath()}\\[2mm]
Returns the full path name and form binary file
  
\item \code{void ToForm::run(Braket &exp, bool print, bool all, string newidlabel)}\\[2mm]
Simplify expression in FORM. Creates file input for FORM, runs the FORM program and returns the result to file and/or screen.

\code{exp} - Braket expression to be simplified in FORM, the result is written back.

\code{print} - if TRUE prints final result to screen

\code{all} - if TRUE writes all the expression members separately in ouput FORM file, if FALSE only writes the full result together

\code{newidlabel} - label to be used when the option to sum index is active
 
\item \code{bool ToForm::getIndexSum()}\\[2mm]
Returns the state of the \code{indexSum} flag.
 
\item \code{void ToForm::setIndexSum(bool flag)}\\[2mm]
Sets the state of the \code{indexSum} flag .
 
\item \code{void ToForm::setRenumber(bool flag=true)}\\[2mm]
Sets "renumber 1;" in FORM input file. This option is used to renumber index in order to allow further simplifications. However, in big expressions this must be avoid since it increases the computational time in FORM. The best way to use is simplify the expression with FORM with this option unset, and then send a second time to FORM with this option active. By default this option is unset.
 
\item \code{bool ToForm::getRenumberOption()}\\[2mm]
Returns the state of the \code{formRenumber} flag.
 
\item \code{ToForm& ToForm::operator<<(const string &func)}\\[2mm]
Overloads operator for \code{ToForm << func}.
 
\item \code{ToForm& ToForm::operator+(const string &func)}\\[2mm]
Overloads operator for \code{ToForm + func}.
\end{itemize}

%%%%%%%%%%%%%%%%%%%%%%%%%%%%
\paragraph{\code{son.h}}

\begin{itemize}	
\item \code{void setDim(int n)}\\	
\code{int getDim()}\\
\code{void CleanGlobalDecl()}\\
\code{void setVerbosity(Verbosity verb)}\\
\code{Verbosity getVerbosity()}\\
\code{Braket Bop(std::string startid="i")}\\
\code{Braket BopIdnum()}\\[2mm]
The description of these functions was done in \cref{sec:generalfunctions}.

\item \code{bool GroupEven()}\\[2mm]
Returns Group parity.

\item \code{ostream& operator<<(ostream& out, const Verbosity &a)}\\[2mm]
Returns current ostream verbosity level. Returns \code{ostream} for output.			

\item \code{size_t getCurrentRSS()}\\[2mm]
Returns the current resident set size (physical memory use) measured in bytes, or zero if the value cannot be determined on this OS.
	
\item \code{size_t getPeakRSS()}\\[2mm]
Returns the peak (maximum so far) resident set size (physical memory use) measured in bytes, or zero if the value cannot be determined on this OS.

\item \code{void print_process_mem_usage()}\\[2mm]
Prints memory stats (current and peak resident set size) in MB.
\end{itemize}

In addition, we have also implemented \Cpp macros that simplifies the call for the functions; its description was done in \cref{sec:generalfunctions}. Notice that the stringising macro operator \texttt{\#}\code{a} causes the argument to be enclosed in double quotation marks.

\begin{itemize}
\item \code{bb(id)}\\[2mm]
This makes the call \code{DList(0, newIdx(id))}, \code{id} is the index in string format or enclosed in quotation marks.

\item \code{bbt(id)}\\[2mm]
This makes the call \code{DList(1, newIdx(id))}, \code{id} is the index in string format or enclosed in quotation marks.
	
\item \code{b(id)}\\[2mm]
This makes the call \code{DList(0, newIdx(#id))}, \code{id} index does not need to be enclosed in quotation marks.
	
\item \code{bt(id)}\\[2mm]
This makes the call \code{DList(1, newIdx(#id))}, \code{id} index does not need to be enclosed in quotation marks.

\item \codeb{delta(id1,id2)}\\[2mm]
This makes the call \code{DList(2, newIdx(#id1), newIdx(#id2))}; \code{id1} and \codeb{id2} indices do not need to be enclosed in quotation marks.
	
\item \code{identity}\\[2mm]
This macro is a shortcut for the object \code{DList(3, 0)}.
	
\item \codeb{bra(i,a,b)}\\[2mm]
This makes the call \codeb{Braket(i, #a, b, bra)}, \code{a} index does not need to be enclosed in quotation marks.
	
\item \codeb{ket(i,a,b)}\\[2mm]
This makes the call \codeb{Braket(i, #a, b, ket)}, \code{i} index sum, \code{a} constant part without quotation marks, \codeb{b} \code{DList} expression.

\item \codeb{braket(i,a,b)}\\[2mm]
This makes the call \codeb{Braket(i, #a, b, braket)}, \code{i} index sum, \code{a} constant part without quotation marks, \codeb{b} \code{DList} expression
		
\item \codeb{free(i,a,b)}\\[2mm]
This makes the call \codeb{Braket(i, #a, b, none)}, \code{i} index sum, \code{a} constant part without quotation marks, \codeb{b} \code{DList} expression.
	
\item \codeb{Field(a, b, c, d)}\\[2mm]
This makes the call \codeb{FormField(#a, b, c, d)}, \code{a} field name without quotation marks, \codeb{b} number of upper index, \code{c} number of lower index, \code{d} field with/without flavor index and symmetric/asymmetric field, returns field name in string format.
\end{itemize}

%%%%%%%%%%%%%%%%%%%%%%%%%%%%%%%%%%%%%%

\end{document}